\documentclass{article}
\pdfoutput=1
\usepackage{jheppub}
\usepackage{amsmath}

\usepackage{tikz}

\newcommand{\roughly}[1]{\mathrel{\raise.3ex\hbox{$#1$\kern-0.85em\lower1ex\hbox{$\sim$}}}}
\newcommand{\lsim}{\roughly<}

\def\cQ{{\cal Q}}

\def\cJ{{\cal J}}

\def\cL{{\cal L}}
\def\cM{{\cal M}}

\def\cO{{\cal O}}
\def\cP{{\cal P}}

\def\cW{{\cal W}}

\newbox\charbox
\newbox\slabox
\def\slsh#1{{      
        \setbox\charbox=\hbox{$#1$}
        \setbox\slabox=\hbox{$/$}
        \dimen\charbox=\ht\slabox
        \advance\dimen\charbox by -\dp\slabox
        \advance\dimen\charbox by -\ht\charbox
        \advance\dimen\charbox by \dp\charbox
        \divide\dimen\charbox by 2
        \raise-\dimen\charbox\hbox to \wd\charbox{\hss/\hss}
        \llap{$#1$}
}}

\def\exd{{\hbox{d}}}

\def\bfx{{\bf x}}
\def\bfy{{\bf y}}
\def\bfz{{\bf z}}
\def\bfr{{\bf r}}
\def\bfe{{\bf e}}
\def\bfE{{\bf E}}
\def\bfA{{\bf A}}
\def\bfB{{\bf B}}

\def\bfJ{{\bf J}}

\def\bfn{{\bf n}}

\def\cQ{{\cal Q}}

\def\cJ{{\cal J}}

\def\cL{{\cal L}}
\def\cM{{\cal M}}
\def\cO{{\cal O}}

\def\cW{{\cal W}}
\def\cP{{\cal P}}

\def\mff{\mathfrak{f}}
\def\mfg{\mathfrak{g}}

\def\bea{\begin{eqnarray}}
\def\eea{\end{eqnarray}}
\def\be{\begin{equation}}
\def\ee{\end{equation}}

\def\ssB{{\scriptscriptstyle B}}

\def\ssD{{\scriptscriptstyle D}}
\def\ssE{{\scriptscriptstyle E}}

\def\ssL{{\scriptscriptstyle L}}

\def\ssN{{\scriptscriptstyle N}}

\def\ssR{{\scriptscriptstyle R}}

\def\KG{{\scriptscriptstyle KG}}

\def\psibar{\overline{\psi}}

\def\Dsl{\slsh{D}}

\def\nn{\nonumber}

\def\({\left(}
\def\){\right)}

\def\pref#1{(\ref{#1})}

\title{Point-Particle Effective Field Theory III: \\Relativistic Fermions and the Dirac Equation}

\author{C.P.~Burgess,}
\author{Peter Hayman,}
\author{Markus Rummel}
\author{and L\'aszl\'o Zalav\'ari}
\affiliation{Department of Physics \& Astronomy, McMaster University\\ $\phantom{wd}$ 1280 Main Street West, Hamilton Ontario, Canada L8S 4M1}
\vspace{2mm}
\affiliation{Perimeter Institute for Theoretical Physics\\  $\phantom{wd}$ 31 Caroline Street North, Waterloo Ontario, Canada  N2L 2Y5 }

\date{\today}

\abstract {We formulate point-particle effective field theory (PPEFT) for relativistic spin-half fermions interacting with a massive, charged finite-sized source using a first-quantized effective field theory for the heavy compact object and a second-quantized language for the lighter fermion with which it interacts. This description shows how to determine the near-source boundary condition for the Dirac field in terms of the relevant physical properties of the source, and reduces to the standard choices in the limit of a point source. Using a first-quantized effective description is appropriate when the compact object is sufficiently heavy, and is simpler than (though equivalent to) the effective theory that treats the compact source in a second-quantized way. As an application we use the PPEFT to parameterize the leading energy shift for the bound energy levels due to finite-sized source effects in a model-independent way, allowing these effects to be fit in precision measurements. Besides capturing finite-source-size effects, the PPEFT treatment also efficiently captures how other short-distance source interactions can shift bound-state energy levels, such as due to vacuum polarization (through the Uehling potential) or strong interactions for Coulomb bound states of hadrons, or any hypothetical new short-range forces sourced by nuclei.}

\begin{document}
\maketitle
\section{Introduction}
\label{section:intro}

Nature is full of examples where small but massive compact objects (of linear size $R$) interact with and control the motions of lighter neighbours within a much larger surrounding domain (of size $a \gg R$). Examples include nuclei and atoms, stars and solar systems as well as legions of others. For such systems familiar arguments (such as the multipole expansion) show that only a few features of the compact object are often relevant to understanding motions in their larger environment. This simplicity usually emerges once observables are expanded in powers of small ratios like $R/a$. 

Effective field theories \cite{EFTs, EFTrevs} are the natural language for exploiting this kind of simplicity, though these are usually only formulated in a second-quantized language with all species of particles represented by their respective quantum field. For instance two-body contact interactions between two species of particles in a fully second-quantized framework would be represented in terms of their respective fields by terms like $g (\psi^*\psi)(\chi^*\chi)$ in an effective Lagrangian. 

Our companion papers \cite{PPEFT1, PPEFT2} explore how to formulate such effective theories using instead a first-quantized language for the heavy compact object, reserving the second-quantized language for the lighter particles with which it interacts.\footnote{Similar methods have been developed to handle compact gravitating systems, such as for gravitational-wave emission by inspiralling compact objects \cite{IraWalter} and gravitational back-reaction in extra-dimensional models to \cite{BraneEFT}.} In this mixed first-quantized/second-quantized (one-two) language, if the heavy ($\chi$) particle is in a position eigenstate situated at $\bfx = 0$ then the two-body contact interaction mentioned above instead has the form $g (\psi^*\psi) \, \delta^3(x)$. This kind of formulation would be appropriate when the mass of the compact object is sufficiently large. In such situations all information about the source enters observables through the boundary conditions that are implied for the light fields at the position of the heavy compact object; boundary conditions that are completely determined by the source's first-quantized effective action. 

This type of one-two formulation can have several advantages. One of these is the more direct connection it provides to the study of particle motion within a central ({\em e.g.} Coulomb or gravitational) potential, for which many useful tools are known (particularly for bound states). In this they are complementary to a fully second-quantized (two-two) formulation, such as for NRQED or NRQCD \cite{NRQECD, EFTscat}, in which induced quantities --- like the nuclear Coulomb potential or solar gravitational field --- arise as a resummation of a particular class of interactions that dominate in some limits. By contrast, in the mixed one-two framework such classical fields are included into the zeroth order description about which one perturbs. 

Furthermore, relating the near-source boundary conditions to the source action takes the guesswork out of small-$r$ boundary conditions, and shows in particular why linear `Robin' boundary conditions are so generic at low energies (see also \cite{Case}). More generally, they show how to handle singular potentials (like $V(r) \propto r^p$ with $p \le -2$) unambiguously, despite the generic absence in these cases \cite{EssinGriffiths} of smooth solutions at the origin.

The study in \cite{PPEFT1, PPEFT2} considered both nonrelativistic and spinless relativistic particles orbiting the massive compact object, focussing in particular on unusual effects that arise if the compact source size, $R$, is small enough that relativistic kinematics is relevant for the matching problem to the interior physics of the source {\em even for bound states whose total energy, $\omega$, is nonrelativistic}: $m - \omega \ll m$. This mixed relativistic/nonrelativistic regime occurs when $mR \ll v \ll 1$, where $v \sim Z \alpha$ is the speed of the orbiting particle (whose mass is $m$). (Here we take the source charge to be $Ze$ and $\alpha = e^2/4\pi$ is the usual fine-structure constant.) 

In particular, for relativistic spinless particles an interesting regime was identified for which energy shifts of $S$-wave states due to the source's finite size scale as 
\be \label{KGscaling}
   \delta \omega_{\KG} \sim \frac{(Z\alpha)^2 R}{m} \left( \frac{mZ\alpha}{n} \right)^3  \propto (Z\alpha)^5 m^2R \,,
\ee
where the last factor is the $S$-wave Schr\"odinger-Coulomb wave-function at the origin $|\psi(0)|^2 \propto (mZ\alpha/n)^3$. Effects like this, scaling linearly with $R$, are unusual and so lead to the question of whether similar shifts occur for the spin-half electrons and muons that arise in conventional and muonic atoms. 

We here address this question by extending the discussion of \cite{PPEFT1, PPEFT2} to spin-half systems, finding that although many of the features of the Klein-Gordon problem of \cite{PPEFT2} also carry over to the Dirac field studied here  the scaling of \pref{KGscaling} does not: the corresponding leading Dirac expression instead gives the standard result: 
\be \label{Dscaling}
 \delta \omega_\ssD \sim  Z\alpha R^2 \left( \frac{mZ\alpha}{n} \right)^3  \propto(Z\alpha)^4  m^3R^2 \,. 
\ee

At first sight this difference in scaling may seem surprising, since  spin-dependent effects in orbital energies might be expected to be suppressed by $v \sim Z\alpha$ leading one to expect Dirac and Klein-Gordon predictions to agree at leading order in $Z\alpha$. Although this expectation is true for most observables, it proves not to be true when tracking finite-size effects because relativistic effects are not small at radii $r \sim R$ once $R \lsim Z\alpha/m \sim (Z\alpha)^2 a_\ssB$ (where $a_\ssB$ is the Bohr radius). Indeed the ratio of $\delta \omega_\KG$ and $\delta \omega_\ssD$ given above is of order $Z\alpha/mR$, which is order unity for electrons (for which $mR \sim Z\alpha$ even though both are separately small).

Along the way we show how to formulate the near-source boundary condition for fermions, and why these differ from those that arise for bosons. We identify how the couplings for two-body contact interactions run, even at the classical level, and how this running goes over to the running found in \cite{PPEFT1,PPEFT2} in the non-relativistic limit. This running properly captures how effective theories can sometimes generate scattering lengths that are much larger than the size $R$ of the underlying object, and corresponds to the first-quantized version of a similar discussion found in \cite{EFTscat}. 

Another result from \cite{PPEFT1,PPEFT2} carries over to fermions: the fixed point of the running is {\em not} at $c_s = c_v = 0$ for charged sources (for which $Z\alpha \ne 0$). It turns out this nontrivial fixed point is precisely what is required in order for the fixed point to reproduce standard results for the Dirac equation in the presence of specific nuclear charge distributions. That is, when we compare the PPEFT approach to explicit solutions to the Dirac equation in the presence of a finite-size charge distribution, we find that matching produces contact interactions for the PPEFT that sit precisely at the infrared fixed point of the RG flow. This shows why energy-level shifts take on a particularly model-independent form (proportional to the charge-radius $r_p^2 = \langle r^2 \rangle$ and higher moments \cite{Friar, Zemach} --- see also \cite{Franziska}) in the special case where the nucleus is modelled as a specific charge distribution. 

In what follows we specialize for simplicity to parity-preserving interactions and spinless compact central objects, and so strictly speaking the interactions we find suffice in themselves to describe finite-size effects in the He${}^+$ ion or muonic states in even-even nuclei \cite{Friar, Muonic, Eides, Puzzle, HillPaz}. The effects we find also apply to nuclei with spin (such as hydrogen) once the effective theory of the first-quantized source is supplemented by the extra interactions that a nuclear spin allows. (We intend to return to discuss spinning sources more fully in a later paper.)

In Section \ref{action_sec} we set the stage by introducing the point-particle effective action in the context of Dirac fermions. In Section \ref{bc_sec} we derive the boundary condition and the induced renormalization group running in the presence and absence of a Coulomb potential. This leads to the discussion of bound state energy shifts implicated by the boundary condition in Section \ref{shifts_sec}. We discuss applications of PPEFT for fermions in Section \ref{examples_sec} and conclude in Section \ref{sum_sec}. We discuss various technicalities in the appendix.

\section{Action and field equations} \label{action_sec}

To make things concrete we focus on describing a relativistic spin-half charged particle interacting with a small charged source. The system of interest consists of a 3+1 dimensional `bulk' action coupled to a 0+1 dimensional `point-particle' action representing the small source ({\em e.g.} the nucleus of an atom),
\be
 S_{\rm tot} = \int \exd^4 x \; \cL_\ssB + \int_\cW \exd \tau \; L_p = \int \exd^4 x \left[ \cL_\ssB + \int_\cW \exd \tau \;\delta^4(x - y(\tau)) L_p \right] \,,
\ee
where $\cW$ indicates the integration is over the world-line, $y^\mu(\tau)$, of the source.  In the final equality $\cL_\ssB$ and $L_p$ are both regarded as being functions of the bulk fields evaluated at an arbitrary spacetime point, $x^\mu$. $L_p$ is also a function of the `brane-localized' position field, $y^\mu(\tau)$. 

\subsection{Action and field equations} 

Taking the bulk dynamics to be QED with a fermion of charge $-e$, the bulk action becomes
\be
 S_\ssB = - \int \exd^4 x \; \left[ \frac14 \, F_{\mu\nu} F^{\mu\nu} + \psibar (\Dsl + m) \psi  \right]\,,
\ee
with $D_\mu  \psi = (\partial_\mu  + i e A_\mu)\psi$. This should be considered in the spirit of a Wilson action, and so in principle also includes an infinite series of subdominant local terms involving more powers of the fields and their derivatives (whose effects are not important in what follows). 

The point-particle action is similarly given by an expansion in these fields, for which (for a spinless, parity-preserving source) the leading parity-even terms are\footnote{Our metric is mostly plus and our Dirac conventions in rectangular and polar coordinates are given in Appendix \ref{App:DiracConventions}.} 
\be \label{sourceaction}
  S_p = -\int_\cW \exd \tau \; \Bigl[ M - Q \, A_\mu \dot y^\mu +  c_s \, \psibar \,\psi +i c_v \, \psibar \,\gamma_\mu \psi \, \dot y^\mu  - {\tilde h} \, \nabla \cdot \bfE + \cdots \Bigr] \,,
\ee
where the over-dot denotes differentiation with respect to proper time,  the coefficients $c_s$, $c_v$ and $\tilde h$ all have dimension length-squared and the ellipses indicate terms suppressed by more than two powers of length. Notice that terms involving more than two powers of $\psi$ first arise suppressed by a coupling with dimension (length)${}^5$, and so are nominally subdominant to several terms involving only two powers of $\psi$ but more derivatives than those written above.

Specializing to the rest frame for a motionless source, $\dot y^\mu (\tau) = \delta^\mu_0$, with charge $Q = Ze$ the bulk field equations become
\be
 (\Dsl + m) \psi + \cJ = 0 \qquad \hbox{and} \qquad
 \partial_\mu F^{\mu \nu} - ie \;\psibar \gamma^\nu \psi + j^\nu = 0 \,,
\ee
where
\be \label{Jcurrent}
 \cJ := - \frac{\partial L_p}{\partial \psibar} = \Bigl( c_s +i c_v \gamma^0 \Bigr)  \psi \, \delta^3(x)  +\cdots \,,
\ee
and 
\be \label{jnu}
 j^\nu := \frac{\partial L_p}{\partial A_\nu} = Ze \left( 1 + \frac{r_p^2}{6} \, \nabla^2 \right) \, \delta^3(x) \; \delta^\nu_0 \,.
\ee
This last equality trades the parameter $\tilde h$ for the mean-square charge radius: $r_p^2 = \langle r^2 \rangle$ of the source charge distribution using $\tilde h = \frac16\, Ze  \, r_p^2$. 

\subsection{Bulk solutions}

We seek solutions to the bulk equations with a motionless point charge situated at the origin.  The Maxwell equation is straightforwardly solved for the given source by choosing $\bfA = 0$ and electrostatic potential
\be \label{Amform}
A^0 = Ze \left[ \frac{1}{4\pi r} - \frac{r_p^2}{6} \, \delta^3(x) \right] \,.
\ee
Here the first term is the usual homogeneous solution to the Poisson equation, normalized using the boundary condition at small radial distance, $r = \epsilon$, corresponding to nonzero electric flux
\be \label{Ambc}
 \oint_{r=\epsilon} \exd^2\Omega \; \bfn \cdot \bfE = Ze \,.
\ee
This boundary condition can be obtained by integrating the Maxwell equation over small Gaussian pillbox of vanishingly small radius $r = \epsilon$. By contrast, the second term in \pref{Amform} is the particular integral arising when solving div $\bfE = - \nabla^2 A^0 = \frac16\, Ze \, r_p^2 \nabla^2 \delta^3(x)$.  

We wish to repeat the above arguments for the Dirac field, whose field equation is
\bea \label{Dirac}
  0 &=& (\Dsl + m) \psi + \left( c_s + ic_v \gamma^0 \right)  \psi \, \delta^3(x) \nn\\
  &=& \left[ -i \gamma^0 \left( \omega + \frac{Z\alpha}{r} \right) + \vec\gamma \cdot \nabla + m \right] \psi + \left( c_s + ic_{v\,{\rm tot}} \gamma^0 \right)  \psi \, \delta^3(x) \,,
\eea
where the second line specializes to energy eigenstates,\footnote{Speaking of `energy eigenstates' for a relativistic field is shorthand for evaluating matrix elements of the form $\langle 0 | \psi(x) | n \rangle$, between the vacuum and an energy eigenstate. The energy $\omega$ is the energy of $|n\rangle$ (relative to the vacuum) and can be found in the usual way from the poles in the correlation functions like $\langle \psibar(x) \psi(y) \rangle$.} $\psi(t) = \psi \, e^{-i\omega t}$, and to gauge potentials of the form \pref{Amform}. The parameter $c_{v\,{\rm tot}}$ denotes the total localized combination
\be
  c_{v\,{\rm tot}} := c_v + \frac{Ze^2}6 \, r_p^2 = c_v + \frac{2\pi}3 \, Z \alpha \, r_p^2 \,.
\ee
This implies $\psi_\ssL$ and $\psi_\ssR$ are related by
\bea \label{Diraccomponents}
\left[ -\left( \omega + \frac{Z\alpha}{r} \right) -i \sigma_k \partial_k  \right] \psi_\ssR + m\, \psi_\ssL + \left( c_s \, \psi_\ssL + ic_{v\,{\rm tot}} \gamma^0  \psi_\ssR \right)  \, \delta^3(x) &=& 0 \nn\\
 \hbox{and} \quad \left[ -\left( \omega + \frac{Z\alpha}{r} \right) +i \sigma_k \partial_k  \right] \psi_\ssL + m\, \psi_\ssR + \left( c_s \, \psi_\ssR +i c_{v\,{\rm tot}} \gamma^0  \psi_\ssL \right)  \, \delta^3(x)  &=& 0 \,.
\eea

Outside the source these equations become $(\Dsl + m)\psi = 0$ which (see Appendix \ref{App:DiracSolutions} for a summary in the present conventions) for rotationally and parity invariant situations have solutions of the parity-even form
\be \label{bulk+}
 \Psi^+ =  \left( \begin{array}{c}  \psi^+_\ssL \\  \psi^+_\ssR  \end{array}  \right)  =  \left( \begin{array}{c}  f_+(r) \,U^+(\theta,\phi) +i g_+(r) \,U^-(\theta,\phi) \\  f_+(r) \,U^+(\theta,\phi) -i g_+(r) \,U^-(\theta,\phi) \end{array}  \right)  \,,
\ee
and parity-odd form
\be \label{bulk-}
 \Psi^- =  \left( \begin{array}{c}  \psi^-_\ssL \\  \psi^-_\ssR  \end{array}  \right)  =  \left( \begin{array}{c}  f_-(r)\, U^-(\theta,\phi) +i g_-(r)\, U^+(\theta,\phi) \\  f_-(r)\, U^-(\theta,\phi) -i g_-(r)\, U^+(\theta,\phi) \end{array}  \right)  \,.
\ee
Here $U^\pm$ are the spinor harmonics that combine the particle's spin-half with orbital angular momenta $\ell = j \mp \frac12$ to give total angular momentum $j = \frac12, \, \frac32, \cdots$.

The functions $f_\pm(r)$ and $g_\pm(r)$ are found by explicitly solving the radial part of the Dirac equation in the presence of a potential $A_0(r)$. For a Coulomb potential with source charge $Ze$ these radial equations are (see Appendix \ref{App:DiracSolutions} for details)
\be \label{fgpluseqstxt}
  f_+' = \left( m + \omega + \frac{Z\alpha}{r} \right) g_+ \quad \hbox{and} \quad
  g_+' + \frac{2g_+}{r} = \left( m - \omega - \frac{Z \alpha}{r} \right) f_+ \,,
\ee
together with 
\be \label{fgminuseqstxt}
  g_-' = \left( m - \omega - \frac{Z\alpha}{r} \right) f_- \quad \hbox{and} \quad
  f_-' + \frac{2f_-}{r} = \left( m + \omega + \frac{Z \alpha}{r} \right) g_- \,.
\ee

These have as their general solutions 
\bea \label{fACgentxt}
    f_\pm &=& \sqrt{m+\omega}\,  e^{-\rho/2} \rho^{\zeta-1} \left\{ A_\pm \, \cM \left[ \zeta -\frac{Z\alpha \omega}{\kappa}, 2\zeta+1; \rho \right] +  C_\pm \rho^{-2\zeta} \cM \left[ -\zeta -\frac{Z\alpha \omega}{\kappa}, -2\zeta+1; \rho \right] \right. \nn\\
    &&\qquad\qquad\qquad\qquad  -  A_\pm \left(\frac{\zeta-{Z\alpha \omega}/{\kappa}}{K-{Z\alpha m}/{\kappa}}\right) \cM \left[ \zeta -\frac{Z\alpha \omega}{\kappa}+1, 2\zeta+1; \rho \right] \\
    &&\qquad\qquad \qquad\qquad\qquad \left. +C_\pm \left(\frac{\zeta+{Z\alpha \omega}/{\kappa}}{K-{Z\alpha m}/{\kappa}}\right) \rho^{-2\zeta} \cM \left[ -\zeta -\frac{Z\alpha \omega}{\kappa}+1, -2\zeta+1; \rho \right] \right\} \,, \nn
\eea
and
\bea \label{gACgentxt}
    g_\pm &=& -\sqrt{m-\omega}\,  e^{-\rho/2} \rho^{\zeta-1} \left\{ A_\pm \, \cM \left[ \zeta -\frac{Z\alpha \omega}{\kappa},2\zeta+1;\rho \right] + C_\pm \rho^{-2\zeta} \cM \left[ -\zeta -\frac{Z\alpha \omega}{\kappa},-2\zeta+1;\rho\right] \right. \nn\\
    &&\qquad\qquad\qquad\qquad  + \left.  A_\pm \left( \frac{\zeta-{Z\alpha \omega}/{\kappa}}{K-{Z\alpha m}/{\kappa}} \right) \cM \left[\zeta -\frac{Z\alpha \omega}{\kappa}+1,2\zeta+1;\rho\right] \right. \\
    &&\qquad\qquad \qquad\qquad\qquad \left.- C_\pm \left( \frac{\zeta+{Z\alpha \omega}/{\kappa}}{K-{Z\alpha m}/{\kappa}}\right) \rho^{-2\zeta} \cM \left[-\zeta -\frac{Z\alpha \omega}{\kappa}+1,-2\zeta+1;\rho\right] \right\} \,.\nn
\eea
Here $A_\pm$ and $C_\pm$ are integration constants, $\cM[a,b; z] = 1 + (a/b)z + \cdots$ are the standard confluent hypergeometric functions, $\omega$ is the mode energy and $\kappa$ and $\zeta$ are defined by
\be
  \kappa = \sqrt{(m-\omega)(m+\omega)}  \qquad \hbox{and} \qquad \zeta = \sqrt{\left(j+\frac12\right)^2 - (Z\alpha)^2} \,,
\ee
with $\kappa$ real because of our focus on bound states: $m > \omega$. The parity of the solution enters the above formulae only through the parameter $K= \mp(j+\frac12)$ where (perversely) standard conventions match negative (positive) $K$ to parity-even (parity-odd) states.

\section{Fermionic boundary conditions and the point-particle action} \label{bc_sec}

The next step is to formalize the boundary conditions at the surface of a spherical Gaussian pillbox of radius $r = \epsilon$, along the lines of what is done in \pref{Ambc} for the Maxwell field. We now show how these relate the constants $c_s$ and $c_v$ of the source action to the ratios $g_+/f_+$ and $f_-/g_-$ at $r = \epsilon$. These boundary conditions are again obtained from the source action by integrating the equations of motion over the interior of the pillbox using the delta-function. 

\subsection{Source-bulk matching}

That is, given the action
\be \label{DiracBaction}
 S = - \int_P \exd^4 x \Bigl[ \sqrt{-g} \; \psibar (\Dsl + m) \psi + \sqrt{-\hat g} \; \psibar N \psi   \, \delta^3(x) \Bigr]   \,,
\ee
where $\hat g_{ab} = g_{\mu\nu} \partial_a x^\mu \partial_b x^\nu$ is the induced metric on the world-volume of the source and $N = c_s + ic_{v\,{\rm tot}} \gamma^0$ is the Dirac matrix specified by the source action $S_p$. Then the $\psi$ equation of motion is
\be \label{DiracBeom}
 \sqrt{-g} \;  (\Dsl + m) \psi + \sqrt{-\hat g} \; N \psi   \, \delta^3(x) = 0  \,,
\ee
so integrating over the small Gaussian pillbox, $P$, of radius $\epsilon$ centred on the source then gives (in the limit $\epsilon \to 0$ of vanishingly small pillbox)
\be
 \lim_{\epsilon \to 0} \int_{\partial P} \exd^2 x \, \sqrt{-g} \; n_\mu \gamma^\mu \psi = \lim_{\epsilon \to 0} \int \exd \theta \exd \phi \, \epsilon^2 \sin\theta  \;  \gamma^r \,\psi = -  \sqrt{-\hat g}\;N \psi(0) \,.
\ee
Here $n_\mu$ is an outward-pointing unit normal to the pillbox so $n_\mu \exd x^\mu = \exd r$, and the integral of the $m \psi$ term vanishes as $\epsilon \to 0$. Our conventions on gamma-matrices in polar coordinates are given in Appendix \ref{App:DiracConventions}.

For spherically symmetric configurations (in the limit where $\epsilon$ is much smaller than all other scales of interest) 
this implies the boundary condition
\be \label{diracBC0}
  \int_{r = \epsilon} \exd^2 \Omega \;  \left[ \epsilon^2 \gamma^r + \frac{1}{4\pi}\Bigl( c_s + ic_{v\,{\rm tot}} \gamma^0 \Bigr) \right] \psi  = 0 \,.
\ee
Notice this boundary condition is trivially satisfied pretty much anywhere in the absence of a source, for a small enough pillbox. This is because no source means $c_s = c_v = r_p = 0$ and $\psi$ varies slowly enough to be approximately constant across the pillbox. In this case the integral over all directions for $\gamma^r$ on the surface of the pillbox gives zero trivially. 

The boundary condition on the Gaussian pillbox can be written as $\int \exd^2 \Omega \; B_\epsilon \, \psi(\epsilon) = 0$ where
\be
 B_\epsilon :=  \gamma^r  + \hat c_s  + i\hat c_v \gamma^0  
   =  \left( \begin{array}{cc}
 \hat c_s & \hat c_v  -i \sigma^r    \\
  {}  \hat c_v +i  \sigma^r  & \hat c_s \end{array}\right) \,.
\ee
The dimensionless coefficients $\hat c_s = c_s/(4\pi\epsilon^2)$ and $\hat c_v = c_{v\,{\rm tot}}/(4\pi\epsilon^2)$ can be interpreted as the coefficients of a term in a `boundary action' defined on the codimension-one surface of the Gaussian pillbox, 
\be
  S_{\rm bound} = -\int_{\partial P} \exd^3 x \; \overline \psi \, \Bigl( \hat c_s + i\hat c_v \, \gamma^0+\cdots \Bigr) \psi \,.
\ee
The subscript $\epsilon$ on $B_\epsilon$ is meant to emphasize that the constants $\hat c_a$ (and in general also the original couplings $c_i$ themselves) also must carry an implicit $\epsilon$-dependence if physical quantities are to remain unchanged as $\epsilon$ is varied (more about which below). 

To see what these boundary conditions mean we write them out separately for $\psi_\ssL$ and $\psi_\ssR$, leading to
\be \label{2partbcv0}
 - \hat c_s \int_\epsilon \exd^2 \Omega \,   \psi^\pm_\ssL= \int_\epsilon \exd^2\Omega \,  \Bigl(   \hat c_v   -i  \sigma^r  \Bigr) \psi^\pm_\ssR \quad \hbox{and} \quad
     - \int_\epsilon \exd^2\Omega \,  \Bigl(  \hat c_v   +i  \sigma^r \Bigr)\, \psi^\pm_\ssL = \hat c_s  \int_\epsilon \exd^2\Omega \,   \psi^\pm_\ssR  \,. 
\ee
Notice that these can be found from one other by making the replacements $\psi_\ssL \leftrightarrow \psi_\ssR$ together with $(\hat c_v,  \hat c_s) \leftrightarrow (-\hat c_v,   -\hat c_s )$. Acting on bulk solutions \pref{bulk+} and  \pref{bulk-} and evaluating the angular integrations, these give
\be  \label{cscvtofg}
  \hat c_s +   \hat c_v   = \frac{c_s+c_{v\,{\rm tot}}}{4\pi \epsilon^2} = \left( \frac{  g_+ }{f_+} \right)_{r=\epsilon} \qquad  \hbox{and} \qquad
  \hat c_s -   \hat c_v  = \frac{c_s-c_{v\,{\rm tot}}}{4\pi \epsilon^2} = \left( \frac{ f_-}{g_-}  \right)_{r=\epsilon} \,.
\ee

In what follows we determine $c_s(R)$ and $c_v(R)$ from several hypothetical UV completions for the structure of the source of size $R$, and then regard \pref{cscvtofg} as a boundary condition that selects the exterior solution appropriate for the source of interest. This emphasizes that it is only through boundary conditions like \pref{cscvtofg} that the physics of a specific source can influence the exterior solution, and so enter into physical observables.

\subsection{RG evolution}
\label{sec:RG}

The radius of the Gaussian pillbox, $r=\epsilon$, where the boundary condition is not a physical scale, and so must drop out of predictions for observables (unlike the physical size, $R$, of the underlying source, say). In detail, this happens because any explicit $\epsilon$-dependence arising in a calculations of an observable cancels an implicit $\epsilon$-dependence buried within the `bare' quantities $c_s$ and $c_v$. Following the procedure of \cite{PPEFT1, PPEFT2} (which in turn builds on \cite{Brenorm}), we next determine what the $\epsilon$-independence of observables implies for the $\epsilon$-dependence of $c_s$ and $c_v$. 

First we establish what is needed to ensure physical quantities remain independent of $\epsilon$. Boundary conditions like \pref{cscvtofg} affect  observables by determining the ratio of the integration constants that arise when integrating the bulk field equations. For instance, writing the general solutions, \pref{fACgentxt} and \pref{gACgentxt}, to the radial part of the Dirac field equation in the form
\be \label{fgvsAC}
 f_\pm(r) = A_\pm f_{1\pm}(r) + C_\pm f_{2\pm}(r) \qquad \hbox{and} \qquad g_\pm(r) = A_\pm g_{1\pm}(r) + C_\pm g_{2\pm}(r) \,,
\ee
it is the two ratios $C_+/A_+$ and $C_-/A_-$ that are determined by a boundary condition like the specification of $(g_\pm/f_\pm)_{r=\epsilon}$. Energy levels for states of either parity are determined by demanding the resulting value for the appropriate $C/A$ be consistent with what is required for $C/A$ by normalizability of the modes at infinity. Scattering amplitudes are similarly determined by $C/A$. It follows that physical predictions are $\epsilon$-independent if $c_s(\epsilon)$ and $c_v(\epsilon)$ are chosen to ensure $C/A$ is $\epsilon$-independent for both parity choices. 

At some level \pref{cscvtofg} says it all. Rather than reading \pref{cscvtofg} as fixing $f_\pm/g_\pm$ at a specific radius given known values of $c_s$ and $c_v$ we can instead read the equations
\begin{equation} \label{csvmatch}
 c_s(\epsilon) = \left[  \frac{g_+(\epsilon)}{f_+(\epsilon)} + \frac{f_-(\epsilon)}{g_-(\epsilon)} \right] 2\pi \epsilon^2 \quad \hbox{and} \quad
 c_{v\,{\rm tot}}(\epsilon) = \left[  \frac{g_+(\epsilon)}{f_+(\epsilon)} - \frac{f_-(\epsilon)}{g_-(\epsilon)} \right] 2\pi \epsilon^2 \,,
\end{equation}
as giving $c_s(\epsilon)$ and $c_{v\,{\rm tot}}(\epsilon)$ for known functions $f_\pm(r)$ and $g_\pm(r)$. This means that the $\epsilon$-dependence of the right-hand-side of \pref{csvmatch} is simply given by the $r$-dependence of $f_\pm(r)$ and $g_\pm(r)$ using \pref{fgvsAC}, with $r=\epsilon$. Because $C_\pm$ and $A_\pm$ are $r$-independent the above conditions tell us what $c_s$ and $c_{v\,{\rm tot}}$ must do to keep them also $\epsilon$-independent.  

Our greatest interest is when $\epsilon$ is much smaller than the typical scale $a$ of the external problem (such as the Bohr radius, for applications to atoms), and in this limit it suffices to use the leading small-$r$ form of the solutions $f_\pm$ and $g_\pm$ when computing the $\epsilon$-dependence of $c_s$ and $c_{v\,{\rm tot}}$. In this regime solutions are usually well described by power laws, with \pref{fgvsAC} reducing to
\be \label{fgvsACsmallr}
 f_\pm(r) = A_\pm \left( \frac{r}{a} \right)^{\zeta-1} + C_\pm \left( \frac{r}{a} \right)^{-\zeta-1} \qquad \hbox{and} \qquad g_\pm(r) = A_\pm \left( \frac{r}{a} \right)^{\zeta-1} + C_\pm \left( \frac{r}{a} \right)^{-\zeta-1} \,.
\ee
For such solutions the choice of $C_\pm/A_\pm$ controls the precise radius at which one of these solutions dominates the other one, and as a result the RG evolution of the couplings implied by \pref{csvmatch} in this regime describes the cross-over between these two types of evolution.  

\subsubsection{Non-relativistic limit} 
\label{runningnonrel_sec}

We start by examining this running for parity-even states in the nonrelativistic limit, which corresponds to the evolution found in \cite{PPEFT1, PPEFT2} using the Schr\"odinger equation. 

The radial equations for parity-even states are given by \pref{fgpluseqstxt} which imply in the nonrelativistic limit (for which the energy and mass are approximately equal, $\omega \simeq m$, and much larger than all other scales) it follows that $g_+ \simeq f_+'/(2m) \ll f_+$. Using this in the second of eqs.~\pref{fgpluseqstxt} and dropping subdominant terms gives the Schr\"odinger equation (in the presence of a Coulomb potential), with Schr\"odinger field $\varphi(r) = f_+(r)$. 

In this limit the Dirac spinor is approximately given by
\begin{equation}
 \psi \simeq \frac{1}{\sqrt{2}}\begin{pmatrix} \varphi \\ \varphi \end{pmatrix} \,,
\end{equation}
so in the nonrelativistic limit the combination appearing in the source action is 
\be
  c_v \psibar \, \psi + i c_v \psibar \gamma^0 \psi \simeq (c_s + c_v) \varphi^* \varphi =: h \, \varphi^* \varphi \,,
\ee
where $h = c_s + c_v$ is the coupling for the analogous effective Schr\"odinger contact interaction.

Defining the quantity $\lambda := 2m\,h_{\rm tot} = 2m\left( h + \frac{2\pi}3 \, Z\alpha \, r_p^2\right)$, the nonrelativistic limit of the boundary condition \pref{cscvtofg} therefore is
\be \label{cscvtofgNR}
 \lambda = 2m\,h_{\rm tot} = 2m(c_s + c_{v\,{\rm tot}}) = 8\pi m\epsilon^2  \left( \frac{g_+}{f_+} \right)_{r=\epsilon}  \simeq 4\pi \epsilon^2 \left( \frac{\varphi'}{\varphi} \right)_{r=\epsilon} \,,
\ee
in agreement with the boundary condition found for a Schr\"odinger field coupled to a source with Lagrangian density $\cL_p = - h\, \varphi^*\varphi \, \delta^3(x)$ \cite{PPEFT1, PPEFT2}. These references also show that restricting to $s$-wave ($\ell = 0$) configurations and using the small-$r$ asymptotic form $\varphi_1(r) \propto r^{\ell}$ and $\varphi_2(r) \propto r^{-\ell-1}$ implies that for small $\epsilon$ the evolution of $h$ given in \pref{cscvtofgNR} satisfies the differential RG equation
\begin{equation}\label{Schroedingerunning}
 \epsilon \frac{\exd\hat\lambda}{\exd\epsilon}   = \frac{1}{2} \left(1-\hat\lambda^2 \right) \qquad \hbox{where} \qquad
  \hat\lambda := \frac{\lambda}{2\pi\epsilon} + 1 = \frac{mh}{\pi\epsilon} +1 \,,
\end{equation}
in which the last equalities define $\hat\lambda$. 

The evolution of $\hat\lambda$ evidently has two fixed points, at $\hat \lambda_\star = \pm 1$, and these respectively correspond to $\lambda_\star = 0$ and $\lambda_\star = - 4\pi \epsilon$. Comparing with \pref{cscvtofgNR} shows these forms for $\lambda_\star$ are equivalent to having $\varphi(r) \propto r^0$ and $\varphi(r) \propto r^{-1}$ ({\em i.e.} $r^\ell$ and $r^{-\ell-1}$ for $\ell = 0$), showing the crossover described below \pref{fgvsACsmallr}. 

\subsubsection{Relativistic running when $Z\alpha = 0$} \label{runningZalpha0_sec}

A similar story relates the solutions $f$ and $g$ to solutions of the Klein-Gordon equation in the relativistic case, as is most easily seen in the absence of the Coulomb interaction ($Z\alpha = 0$), as we now show.

\subsubsection*{Parity-even case}

When $Z \alpha = 0$ the first of eqs.~\pref{fgpluseqstxt} again gives $g_+$ as the derivative of $f_+$:
\be
  g_+ = \frac{f_+'}{m+\omega} \,,
\ee
for a mode of energy $\omega$. Using this in the second equation then shows $f_+$ satisfies the Klein-Gordon equation. This shows that the $r$-dependence of the ratio $g_+/f_+$ is proportional to the ratio $\chi'/\chi$ for a Klein-Gordon field:
\be
 \left( \frac{g_+}{f_+} \right)_{r=\epsilon} = \frac{1}{m+\omega} \left( \frac{\chi'}{\chi} \right)_{r=\epsilon} \,.
\ee

But refs.~\cite{PPEFT1, PPEFT2} show (even for $Z\alpha \ne 0$) that if we define the quantity 
\be \label{lambdaKGdef}
 \lambda = 4\pi \epsilon^2 \left( \frac{\chi'}{\chi} \right)_{r = \epsilon} \,,
\ee
for $\chi$ a general $\ell = 0$ solution to the Klein-Gordon equation, then $\hat \lambda := (\lambda/2\pi \epsilon)+1$
satisfies the RG equation
\begin{equation}\label{Schroedingerunning2}
 \epsilon \frac{\exd}{\exd\epsilon} \left( \frac{\hat\lambda}{\zeta_s} \right) = \frac{\zeta_s}{2} \left[1-\left(\frac{\hat\lambda}{\zeta_s}\right)^2 \right]
\end{equation}
for $\epsilon$ small enough to use the small-$r$ asymptotic solution for $\chi(r)$. Here $\zeta_s := \sqrt{1 - 4 (Z\alpha)^2}$. As $Z\alpha \to 0$ it follows $\lambda$ as defined in \pref{lambdaKGdef} again satisfies the RG equation \pref{Schroedingerunning}. 

These considerations show that when $Z\alpha $ vanishes, if we define the quantity
\be
 \lambda_\ssD^+ := (m+\omega) (c_s + c_v) = (m+\omega) 4\pi \epsilon^2 \left( \frac{g_+}{f_+} \right) = 4\pi \epsilon^2 \left( \frac{\chi'}{\chi} \right) \,,
\ee
for parity-even $j = \frac12$ states, then $\hat \lambda_\ssD^+ := (\lambda_\ssD^+/2\pi \epsilon) + 1$ satisfies the same RG equation, eq.~\pref{Schroedingerunning}, as does $\hat \lambda$ in the Klein-Gordon case. Notice that in the nonrelativistic limit we have $\lambda_\ssD^+ \to 2m(c_s+c_v)$ in agreement with the $Z\alpha \to 0$ limit of \pref{cscvtofgNR}. 

\subsubsection*{Parity-odd case}

A similar argument goes through for the parity-odd $j=\frac12$ states. Parity-odd states satisfy the radial equations \pref{fgminuseqstxt} and so when $Z\alpha = 0$ we have
\be
  f_- = \frac{g_-'}{m-\omega} \,.
\ee
Repeating the arguments of the parity-odd case then shows that $g_- = \chi$ satisfies the Klein-Gordon equation and so implies that $\hat \lambda^-_\ssD = (\lambda^-_\ssD/2\pi \epsilon)+1$ satisfies (for small $\epsilon$) the same RG equation, \pref{Schroedingerunning} as do the parity-even and Klein-Gordon cases, provided we define
\be
 \lambda_\ssD^- := (m-\omega) (c_s - c_v) = (m-\omega) 4\pi \epsilon^2 \left( \frac{f_-}{g_-} \right) = 4\pi \epsilon^2 \left( \frac{\chi'}{\chi} \right) \,.
\ee

\subsubsection*{Flow patterns}

\begin{figure}[h]
\begin{center}
 \includegraphics[width=0.8\textwidth]{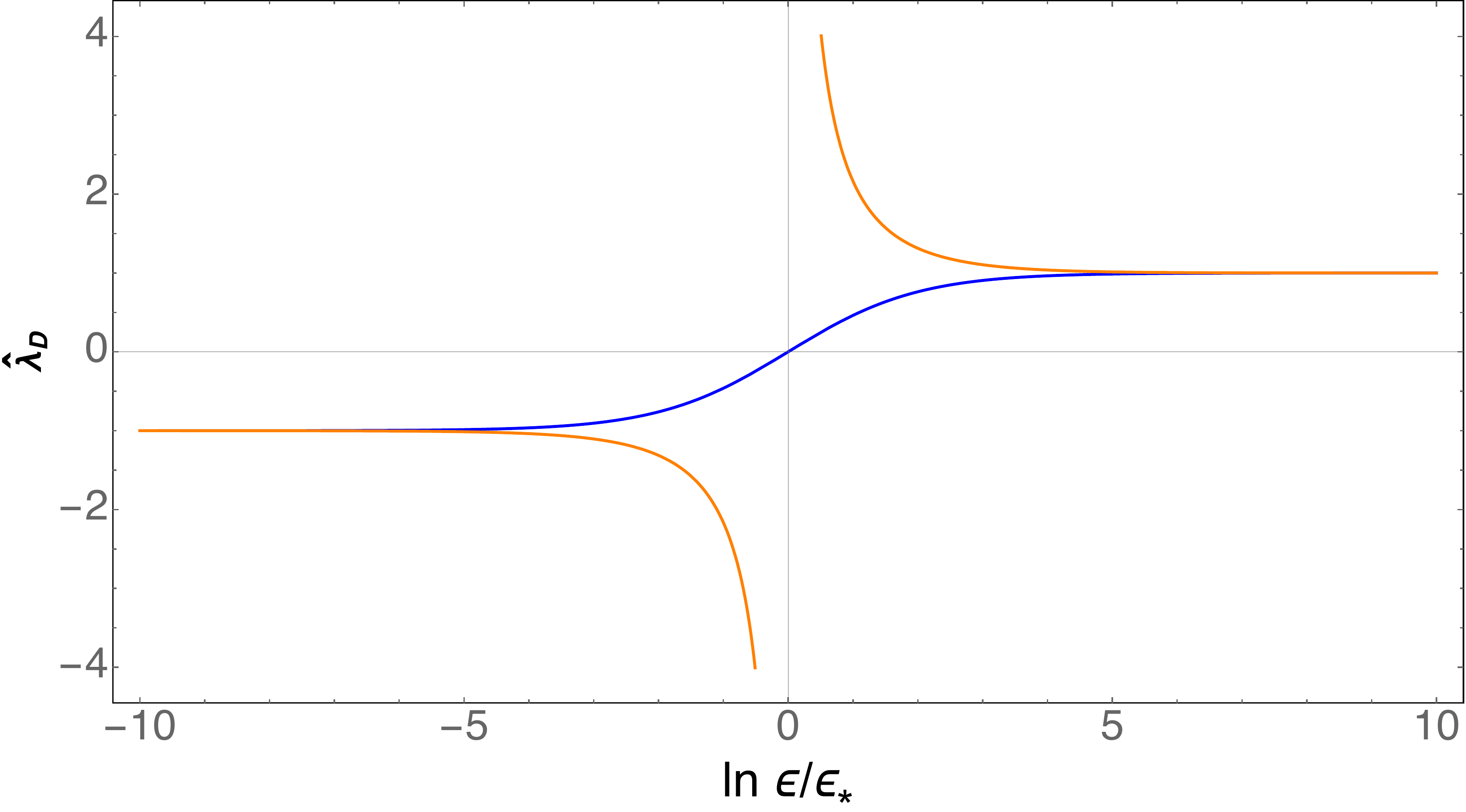} 
\caption{Plot of the RG flow of $\hat\lambda_\ssD^\pm$ (as defined in the main text) vs $\ln \epsilon/\epsilon_\star$ when $Z\alpha = 0$. A representative of each of the two RG-invariant classes of flows is shown, and $\epsilon_\star$ is chosen as the place where $\hat \lambda = 0$ or $\hat \lambda \to \infty$, depending on which class of flows is of interest.} \label{fig:RGflow} 
\end{center}
\end{figure}

The flow obtained by integrating \pref{Schroedingerunning} is given (for $\epsilon$ small enough that $f$ and $g$ are dominated by their near-source asymptotic forms) by
\be \label{lambdaeps}
 \hat \lambda_\ssD^\pm(\epsilon) 
 =  \frac{ \hat \lambda_0^\pm (\epsilon + \epsilon_{0\pm})+ (\epsilon - \epsilon_{0\pm})}{(\epsilon + \epsilon_{0\pm})+ \hat \lambda_0^\pm (\epsilon - \epsilon_{0\pm})}   = \left( \frac{ \epsilon + \epsilon_{\star\pm}}{\epsilon - \epsilon_{\star\pm}} \right)^{\eta_\pm}   \,,
\ee
a flow that is shown in Fig.~\ref{fig:RGflow}. In the first equality the integration constant is chosen using the initial condition $\lambda_\ssD^\pm(\epsilon_{0\pm}) = \lambda^\pm_0$, while in the second equality $\eta_\pm = \hbox{sign}(|\hat\lambda^\pm_\ssD|-1)$ and the RG-invariant quantities $\epsilon_{\star\pm}$ are defined as the scales where the $\hat \lambda_\ssD^\pm$ approach zero (or diverge). Which of these one uses depends on whether the RG trajectory of interest has $|\hat\lambda_\ssD^\pm|$ greater than or smaller than 1. In either case $\epsilon_{\star\pm}$ is given explicitly by inverting the first equality of \pref{lambdaeps}:
\be \label{lambdaepsinv}
 \frac{\epsilon_{\star\pm}}{\epsilon_{0\pm}}  =  \lim_{\lambda^\pm_\ssD \to {0 \atop \infty}} \frac{ \hat \lambda_\ssD^\pm \hat \lambda^\pm_0 - 1 - (\hat \lambda_\ssD^\pm - \hat \lambda^\pm_0)  }{ \hat \lambda_\ssD^\pm \hat \lambda_0^\pm -1 + (\hat \lambda_\ssD^\pm - \hat \lambda^\pm_0)  } = \eta_\pm \, \left(
 \frac{\hat\lambda_0^\pm-1}{\hat\lambda_0^\pm+1} \right) \,.
\ee
As shown in detail in \cite{PPEFT1, PPEFT2}, the $\epsilon$-independence of physical quantities implies they depend only on $\lambda_\ssD^\pm(\epsilon)$ and $\epsilon$ through RG-invariant quantities like $\epsilon_{\star\pm}$.  

For $\epsilon\gg \epsilon_{\star\pm}$ (though $\epsilon$ not so large as to invalidate the small-$r$ expansion of the mode functions at $r=\epsilon$) the flow approaches the fixed point at $\hat \lambda_\ssD^\pm = +1$, with $\hat \lambda_\ssD^\pm -1 \propto \epsilon_{\star\pm}/\epsilon$. Because $\hat \lambda_\ssD^\pm -1 \propto (c_s\pm c_v)/\epsilon$ this implies $c_s$ and $c_v$ simply become independent of $\epsilon$ in this limit.

For small $\epsilon$ the flow emerges from the repulsive fixed point at $\hat \lambda^\pm_\ssD = -1$ with $\hat\lambda_\ssD^\pm +1 \simeq -2\eta_\pm (\epsilon/\epsilon_{\star\pm})$ with (as before) $\eta_\pm = \hbox{sign}(|\hat\lambda^\pm_\ssD|-1)$.  Consequently for small $\epsilon$ the couplings $c_s$ and $c_v$ evolve {\em linearly} with $\epsilon$ (as opposed to the naive quadratic behaviour expected on dimensional grounds):
\bea
  c_s(\epsilon) &=& \frac12 \left( \frac{\lambda^+_\ssD}{m+\omega} + \frac{\lambda^-_\ssD}{m-\omega} \right) =- \frac{4\pi m \epsilon}{m^2 - \omega^2} + \cO(\epsilon^2) \nn\\
  \hbox{and } \qquad
  c_v(\epsilon) &=& \frac12 \left( \frac{\lambda^+_\ssD}{m+\omega} - \frac{\lambda^-_\ssD}{m-\omega} \right) =  \frac{4\pi \omega \epsilon}{m^2 - \omega^2} + \cO(\epsilon^2) \,.
\eea
The flow describes the transition between these two asymptotic states, and clearly no source coupling ($c_s = c_v = 0$) is an RG-invariant fixed point, and it is also RG-invariant to have $c_v = 0$ while $c_s$ runs (corresponding to $\epsilon_{\star+}=\epsilon_{\star-}$). 

As a concrete example, suppose matching to a UV completion were to give the predictions 
\be
  c_v = g_v R^2 \qquad \hbox{and} \qquad c_s = g_s R^2 \qquad \hbox{at $\epsilon = R$,}
\ee
for a microscopic scale $1/R \gg \omega \ge m$ and dimensionless constants $|g_v| ,\,|g_s| \lsim \cO(1)$ . Then $\lambda_\ssD^\pm(R) = (m \pm \omega) (g_s \pm g_v) R^2$ while the signs $\eta_\pm = \hbox{sign}(\hat \lambda^\pm_\ssD-1)$ are $\eta_+ =  \hbox{sign} (g_s + g_v)$ and $\eta_- =  \hbox{sign}\left(g_v - g_s \right)$. Then the RG-invariant scales are $\epsilon_{\star\pm} /R = \eta_\pm (\hat \lambda_\ssD^\pm -1)/(\hat \lambda_\ssD^\pm + 1)$ and so
\be
 \frac{\epsilon_{\star\pm}}{R} =  \eta_\pm \left[ \frac{(m\pm\omega)(g_s\pm g_v)R/4\pi}{1+(m\pm\omega)(g_s\pm g_v)R/4\pi} \right]  
  \,,
\ee
and so $\epsilon_{\star\pm} \gg R$ requires $(g_s \pm g_v)R \simeq -4\pi/(m\pm \omega)$. Unlike for the nonrelativistic case there is always an $\omega$ for which this can be satisfied, but because $\omega R \ll 1$ this is only possible in the effective theory if $g_s \pm g_v$ is sufficiently large and has the right sign. 

For general $\epsilon$ the running couplings are
\be
 \hat \lambda_\ssD^\pm (\epsilon) = \left( \frac{ \epsilon + \epsilon_{\star\pm}}{\epsilon - \epsilon_{\star\pm}} \right)^{\eta_\pm} =  \frac{ \epsilon + (\epsilon+ R)(m\pm\omega)(g_s\pm g_v)R/4\pi}{ \epsilon + (\epsilon-R)(m\pm\omega)(g_s\pm g_v)R/4\pi}  \,,
\ee
which has the right limits for both large and small $\epsilon$. Consequently
\be
  c_s(\epsilon) \pm c_v(\epsilon)   = \frac{2\pi \epsilon}{m\pm\omega} \Bigl( \hat \lambda_\ssD^\pm - 1 \Bigr)  =  \frac{ (g_s\pm g_v)  R^2 }{ 1 + (1- R/\epsilon)(m\pm\omega)(g_s\pm g_v)R/4\pi} \,,
\ee
which shows how the flow for $\epsilon \gg \epsilon_{\star\pm}$ is towards constant $c_s$ and $c_v$, asymptoting to limits renormalized relative their values at $\epsilon = R$. 

\subsubsection{Relativistic running when $Z\alpha \neq 0$}

We repeat the analysis of Section \ref{runningZalpha0_sec} this time for the case $Z\alpha \neq 0$ as is relevant to the Coulomb problem. 

\subsubsection*{Parity even}

The running in the parity even case is determined by equation \eqref{cscvtofg}. The small radius expansion of the mode functions $f$ \eqref{fACgentxt} and $g$ \eqref{gACgentxt} yields to leading order
\begin{equation} \label{cs+cvsmallrhoexp}
	\hat c_s + \hat c_v = \left( \frac{  g_+ }{f_+} \right)_{r=\epsilon} \simeq -\sqrt{\frac{m-\omega}{m+\omega}}\,\frac{\left[(1-\zeta)\kappa + (m+\omega)Z\alpha \right] \left(2\kappa \epsilon \right)^{2\zeta} + \left[(1+\zeta) \kappa + (m+\omega)Z\alpha \right] \frac{C_+}{A_+}}{\left[(1+\zeta) \kappa + (m-\omega)Z\alpha \right]\left(2\kappa \epsilon \right)^{2\zeta} + \left[(1-\zeta) \kappa + (m-\omega)Z\alpha \right] \frac{C_+}{A_+}}\,.
\end{equation}
The RG running can be found by calculating the derivative $d (\hat c_s + \hat c_v) / d\epsilon$ and after inverting \eqref{cs+cvsmallrhoexp} inserting $\epsilon^{2\zeta}$ as a function of $\hat c_s + \hat c_v$:
\begin{equation} \label{cs+cvrun}
 \epsilon \,\frac{\exd (\hat c_s + \hat c_v)}{\exd\epsilon} = -Z\alpha\, \left[ \left(\hat c_s + \hat c_v + \frac{1}{Z\alpha} \right)^2 - \left(\frac{\zeta }{Z\alpha} \right)^2 \right]\,.
\end{equation}

Defining the quantity
\begin{equation}
	\label{eq:def_lamP}	
	\hat\lambda^+_D := Z\alpha(\hat c_s + \hat c_v) + 1,
\end{equation}
the RG equation \eqref{cs+cvrun} takes the form
\begin{equation}
	\label{cs+cvrun2}
	\epsilon\frac{\exd}{\exd\epsilon}\left( \frac{\hat\lambda_\ssD^+}{\zeta} \right) = \zeta \left[ 1 - \left( \frac{\hat\lambda_\ssD^+}{\zeta} \right)^2 \right],
\end{equation}
which has the solution
\begin{equation}
	\label{eq:runPlus}
	\frac{\hat\lambda_\ssD^+}{\zeta} = \frac{\hat\lambda^+_{D0}/\zeta + \tanh[\zeta\ln(\epsilon/\epsilon_0)]}{1 + (\hat\lambda^+_{D0}/\zeta)\tanh[\zeta\ln(\epsilon/\epsilon_0)]} = \frac{(\lambda^+_{D0} + \zeta)(\epsilon/\epsilon_0)^{2\zeta} + (\lambda^+_{D0} - \zeta)}{(\lambda^+_{D0} + \zeta)(\epsilon/\epsilon_0)^{2\zeta} - (\lambda^+_{D0} - \zeta)} \,.
\end{equation}

\subsubsection*{Parity odd}

Similarly to the parity even case we can write \eqref{cscvtofg}  as
\begin{equation} \label{cs-cvsmallrhoexp}
	\hat c_s - \hat c_v = \left( \frac{  f_- }{g_-} \right)_{r=\epsilon} \simeq -\sqrt{\frac{m+\omega}{m-\omega}}\,\frac{\left[(1-\zeta) \kappa - (m-\omega)Z\alpha \right] \left(2\kappa \epsilon \right)^{2\zeta} + \left[(1+\zeta) \kappa - (m-\omega)Z\alpha \right] \frac{C_-}{A_-}}{\left[(1+\zeta) \kappa - (m+\omega)Z\alpha \right]\left(2\kappa \epsilon \right)^{2\zeta} + \left[(1-\zeta) \kappa - (m+\omega)Z\alpha \right] \frac{C_-}{A_-}}\,.
\end{equation}
Repeating the procedure of the previous subsection we then find the running to be
\begin{equation} \label{cs-cvrun}
 \epsilon \,\frac{\exd (\hat c_s - \hat c_v)}{\exd\epsilon} = Z\alpha\, \left[ \left(\hat c_s - \hat c_v - \frac{1}{Z\alpha} \right)^2 - \left(\frac{\zeta }{Z\alpha} \right)^2 \right]\,.
\end{equation}

	Again, one can define the quantity
\begin{equation}
	\label{eq:def_lamM}	
	\hat\lambda^-_D := Z\alpha(\hat c_s - \hat c_v) - 1,
\end{equation}
in terms of which the RG equation \eqref{cs-cvrun} takes the form
\begin{equation}
	\label{cs-cvrun2}
	\epsilon\frac{\exd}{\exd\epsilon}\left( \frac{\hat\lambda_\ssD^-}{\zeta} \right) = -\zeta \left[ 1 - \left( \frac{\hat\lambda_\ssD^-}{\zeta} \right)^2 \right],
\end{equation}
which has the solution
\begin{equation}
	\label{eq:runMinus}
	\frac{\hat\lambda_\ssD^-}{\zeta} = \frac{\hat\lambda^-_{D0}/\zeta - \tanh(\zeta\ln(\epsilon/\epsilon_0))}{1 - (\hat\lambda^-_{D0}/\zeta)\tanh(\zeta\ln(\epsilon/\epsilon_0))} = \frac{ (\lambda^-_{D0} + \zeta)+(\lambda^-_{D0} - \zeta)(\epsilon/\epsilon_0)^{2\zeta} }{(\lambda^-_{D0} + \zeta)-(\lambda^-_{D0} - \zeta)(\epsilon/\epsilon_0)^{2\zeta} } \,.
\end{equation}

\subsubsection*{Fixed points}

From the running equations \eqref{cs+cvrun2} and \eqref{cs-cvrun2}, it is clear that there are fixed points when $\hat\lambda_\ssD^+ = \pm \zeta$, and when $\hat\lambda_\ssD^- = \pm \zeta$. However, from the solutions \eqref{eq:runPlus} and \eqref{eq:runMinus}, we see that the fixed points of $\lambda_\ssD^\pm$ are coupled. The fixed point obtained in the limit $\epsilon \to \infty$ (which we call the IR fixed point) corresponds to be $\lambda_\ssD^+ = +\zeta$ and $\lambda_\ssD^- = -\zeta$, so that
\begin{alignat}{2}
	\label{eq:fixedIR}
	\hat c_s &= 0 \quad &\text{and} \quad \hat c_v &= \frac{\zeta - 1}{Z\alpha} \qquad \text{(IR)}.
\end{alignat}
The UV fixed point is similarly defined as the limit $\epsilon \to 0$ and is given by $\lambda_\ssD^+ = -\zeta$ and $\lambda_\ssD^- = +\zeta$, so that
\begin{alignat}{2}
	\label{eq:fixedUV}
	\hat c_s &= 0 \quad &\text{and} \quad \hat c_v &= - \left( \frac{\zeta + 1}{Z\alpha}\right) \qquad \text{(UV)}.
\end{alignat}

For later purposes (when comparing to results for specific nuclear charge distributions) we remark that the IR fixed point implies the couplings $c_s$ evaluate at $\epsilon = R$ to
 \begin{equation} \label{cs+cvIR}
  (c_s+c_{v})_{\scriptscriptstyle IR} \simeq - 2\pi Z\alpha\,R^2\,,
 \end{equation}
which uses $\zeta \simeq 1 - \frac12(Z\alpha)^2$. 

The attentive reader may also be puzzled as to why the running for $Z\alpha \to 0$ does not coincide with the $Z\alpha = 0$ running found earlier. The reason for this is the observation that the limits $\epsilon \to 0$ and $Z \alpha \to 0$ do not commute, due to the appearance of factors of $1/(1-\zeta) \simeq 1/(Z\alpha)^2$ within the hypergeometric functions that furnish the Dirac-Coulomb solutions. (Related to this, mode functions can asymptote to $r^p$ at small $r$ where $p \propto (Z\alpha)^2$, again displaying non-commuting small-$r$ and $Z\alpha \to 0$ limits.) As discussed in later sections, this makes the evaluation of energy shifts for bound states for specific values for $Z\alpha$ and nuclear size $R$ somewhat subtle, since care must be taken to work to a consistent order in small quantities.

\subsection{Higher-order interactions}

For some applications it is insufficient to work only to lowest order in the nuclear size, and so we pause here to classify some of the next-to-leading interactions according to their dimension:
\be 
  S_p = \int \exd^4 x \Bigl[ \cL_0 + \cL_1 + \cL_3 + \cL_4 + \cL_5 + \cdots \Bigr]  \,,
\ee
where the operators appearing in $\cL_n$ has engineering dimension (mass)${}^n$. In this notation $\cL_0 + \cL_1 + \cL_3$ represent the terms already written in \pref{sourceaction}, so we now enumerate the dimension-4 interactions. At this order the operators consistent with invariance under rotations, gauge transformations and C, P and T are $\bfE^2$, $\bfB^2$ and\footnote{A spatial derivative, $\psibar \, \vec \gamma \cdot \nabla \psi$, need not be included separately since it is redundant --- {\em i.e.} it can be recast in terms of one of those already written by a field redefinition and/or an integration by parts.} $\psibar\, \gamma^0 D_0 \psi$.  We therefore take
\be \label{nexdim}
 \cL_4 = - \left[ \frac12 \Bigl( \tilde h_\ssE \, \bfE^2 + \tilde h_\ssB \, \bfB^2 \Bigr) + c_t \, \psibar \, \gamma^0 D_0 \psi \right] \delta^3(x) \,,
\ee
where $c_t$ and the `polarizabilities' $\tilde h_\ssE$ and $\tilde h_\ssB$ are new effective couplings having dimension (length)${}^3$. For instance the time derivative appearing in the last of these terms contains contributions to the Dirac equation that resemble a correction to $c_v$ by an amount $\delta c_v \propto c_t \omega$. For nonrelativistic bound states and for $c_v \propto R^2$ and $c_t \propto R^3$ such corrections look like $mR^3|\phi(0)|^2$ contributions to the energy shift, and so contribute to some of the subleading corrections discussed below.

One can continue in this way to as high a dimension as one wishes. Notice that the first interaction to involve more than two Dirac fields --- such as `three-body' interactions, like $c_{3b} \, (\psibar\,\psi) \, (\psibar\,\psi) \; \delta^3(x)$ --- arises once we consider effective couplings with dimension (length)${}^5$.

Effectively, we can parametrize the boundary condition as
\be \label{xig}
 \left( \frac{g_+}{f_+} \right)_{r=R} = \xi_g\,  Z \alpha \qquad \hbox{with} \qquad
 \xi_g = \hat g_1 + \hat g_2 (mRZ\alpha) + \hat g_3 (Z\alpha)^2 + \cdots \,.
\ee
Any microscopic source physics can only influence parity-even physical observables through their contributions to the constants $\hat g_i$, only a few of which are relevant to any given order in the small expansion parameters. This makes these parameters useful proxies for specific models of source physics, and their values are computed in Appendix \ref{App:DiracSolutions} for several simple examples. Although quantities like $\hat g_2$ can be traded for parameters like $c_t$ and/or $h_\ssE$ we do not pursue this connection explicitly here.

\section{Bound-state energy shifts} \label{shifts_sec}

With a view to computing nuclear-size effects on atomic energy levels we next turn to the implications source contact interactions have for the energy of states bound to the source. Our assumptions of rotation invariance in $S_p$ restricts us for simplicity to atoms with spherically symmetric nuclei. What we find also applies to nuclei with spin but must be supplemented by spin-dependent nuclear-size effects (such as nuclear-size effects for hyperfine splitting \cite{Zemach}).

\subsection{Energy-shift calculations}

Bound-state energies are computed by reconciling the implications for the integration constants, $C_\pm/A_\pm$, appearing in \pref{fgvsAC} (or in more detail \pref{fACgentxt} and \pref{gACgentxt}) as imposed by the small-$r$ and large-$r$ boundary conditions. At small $r$ the relevant boundary conditions are \pref{cscvtofg}, which we repeat here for convenience
\be  \label{cscvtofg2}
  \hat c_s +   \hat c_v   = \frac{c_s+c_{v\,{\rm tot}}}{4\pi \epsilon^2} = \left( \frac{  g_+ }{f_+} \right)_{r=\epsilon} \qquad  \hbox{and} \qquad
  \hat c_s -   \hat c_v  = \frac{c_s-c_{v\,{\rm tot}}}{4\pi \epsilon^2} = \left( \frac{ f_-}{g_-}  \right)_{r=\epsilon} \,,
\ee
and the implications of these for $C_\pm/A_\pm$ --- as found using \pref{fACgentxt} and \pref{gACgentxt} --- must be consistent with normalizability at large $r$, which implies
\begin{equation} \label{ACGammastxt}
 - \, \frac{C_\pm}{A_\pm} = \frac{\Gamma(1+2\zeta)}{\Gamma(1-2\zeta)}\,\frac{\Gamma(-\zeta-{Z\alpha \omega}/{\kappa})}{\Gamma(\zeta-{Z\alpha \omega}/{\kappa})} \,.
\end{equation}

\begin{figure}[h]
\begin{center}
\includegraphics[width=0.8\textwidth]{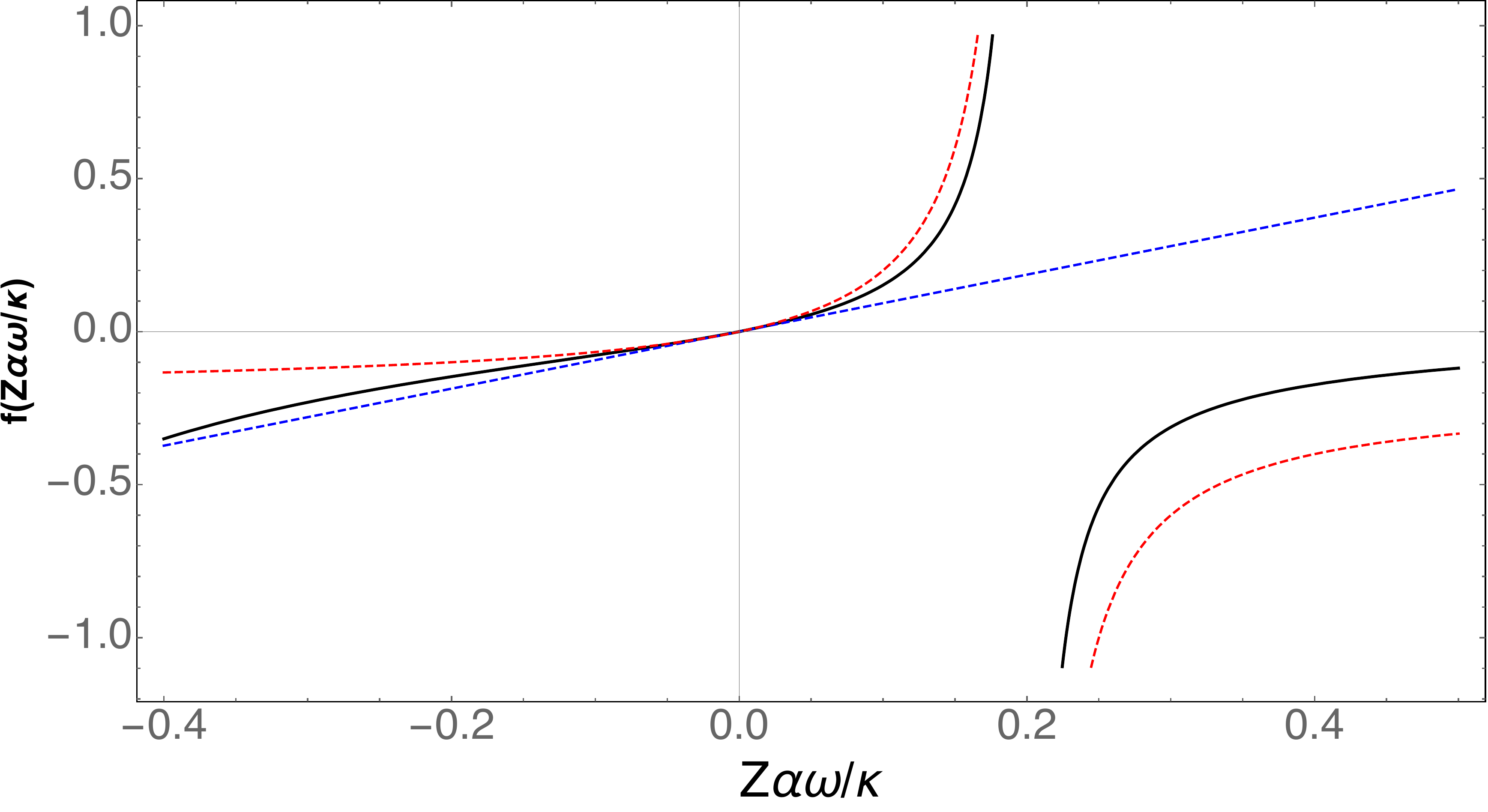}
\caption{\small The black curve plots the right-hand side of eq.~\pref{ACGammastxt} $f(Z\alpha \omega/\kappa) = \Gamma(1+2\zeta) \Gamma(-\zeta-{Z\alpha \omega}/{\kappa}) / ( \Gamma(1-2\zeta) \Gamma(\zeta-{Z\alpha \omega}/{\kappa}))$ vs $Z\alpha \omega/\kappa$, with the zero of energy chosen to be the eigenvalue of the $n=2$ and $j=\frac12$ states. Standard Dirac energy levels correspond to places where the plotted quantity vanishes, while finite-size effects of the source correspond to those energies for which \pref{ACGammastxt} instead equals a specified nonzero (positive) value. The dashed curves show two approximations to \pref{ACGammastxt} that provide useful analytic expressions for energy shifts. The  blue (red) curve shows the single-pole (double-pole) approximation to \pref{ACGammastxt}, described in the main text.  In order to better display the shape of these curves, for plotting purposes we use $\zeta = 0.9$ (and so $Z\alpha \sim 0.45$) and for concreteness expand about the pole at $n=2$.  }\label{Fig:DiracPoleCheck}
\end{center}
\end{figure}

In the absence of a source the Dirac energy eigenvalues are given by solutions to $C_\pm/A_\pm = 0$, which \pref{ACGammastxt} shows is satisfied when $\zeta - Z\alpha \omega/\kappa = -N$ with $N = 0,1,2,\cdots$. This returns the standard Dirac energy eigenvalues
\be \label{DiracE0txt}
 \omega_{\ssN} =  m \left[ 1 + \frac{(Z \alpha)^2}{ \left(n + \zeta - j - \frac12 \right)^2} \right]^{-1/2} \simeq m \left[ 1 - \frac{(Z\alpha)^2}{2n^2} - \frac{[4n-3(j+1/2)]}{8n^4(j+1/2)} \, (Z\alpha)^4+\cO[(Z\alpha)^6 \right]  \,,
\ee
where $n = N + \left( j + \frac12 \right) = 1,2,3, \cdots$ is the usual principal quantum number. 

In the presence of a finite-sized source we instead solve for $\omega$ by equating the right-hand side of \pref{ACGammastxt} to the nonzero value of $C/A$ obtained by fixing $f/g$ using the boundary condition \pref{cscvtofg} at nonzero $r=\epsilon$. In practice this is done in two steps: ($i$) computing the value of $C/A$ implied from the microscopic physics of the source (as parametrized by $S_p$, say); and ($ii$) solving \pref{ACGammastxt} for $\omega$ as a function of nonzero $C/A$, given a known form for $C/A$. We next consider each of these steps in turn.

\subsubsection*{Solving for $\delta \omega$}

Solving for $\delta \omega = \omega - \omega_{\ssN}$ with given $C/A$ requires no knowledge of source structure since the right-hand side of \pref{ACGammastxt} is dictated purely by the known solutions to the Coulomb-Dirac equation. Although this is easily done numerically, there are also accurate analytic approximations that are very useful (particularly when tracking the dependence of the result on external parameters), which are summarized briefly here.

Figure \ref{Fig:DiracPoleCheck} plots the right-hand side of \pref{ACGammastxt} against energy with the zero of energy chosen to be the Dirac energy eigenvalue for a point-like source corresponding to a particular whole number $N$. Also plotted are two approximate forms, corresponding to approximating $\Gamma(-N+\delta z) \simeq (-)^N/[N! \, \delta z]$ in just the denominator ({\em single-pole approximation}) or in both the denominator and numerator ({\em double-pole approximation}). As the figure shows, because of the presence of a nearby pole in the numerator the first of these approximations turns out only to have a radius of convergence of order $(1-\zeta) \sim (Z\alpha)^2$ and so is only of use for extremely small $\delta z$.

\begin{figure}[h]
\begin{center}
\includegraphics[width=0.8\textwidth]{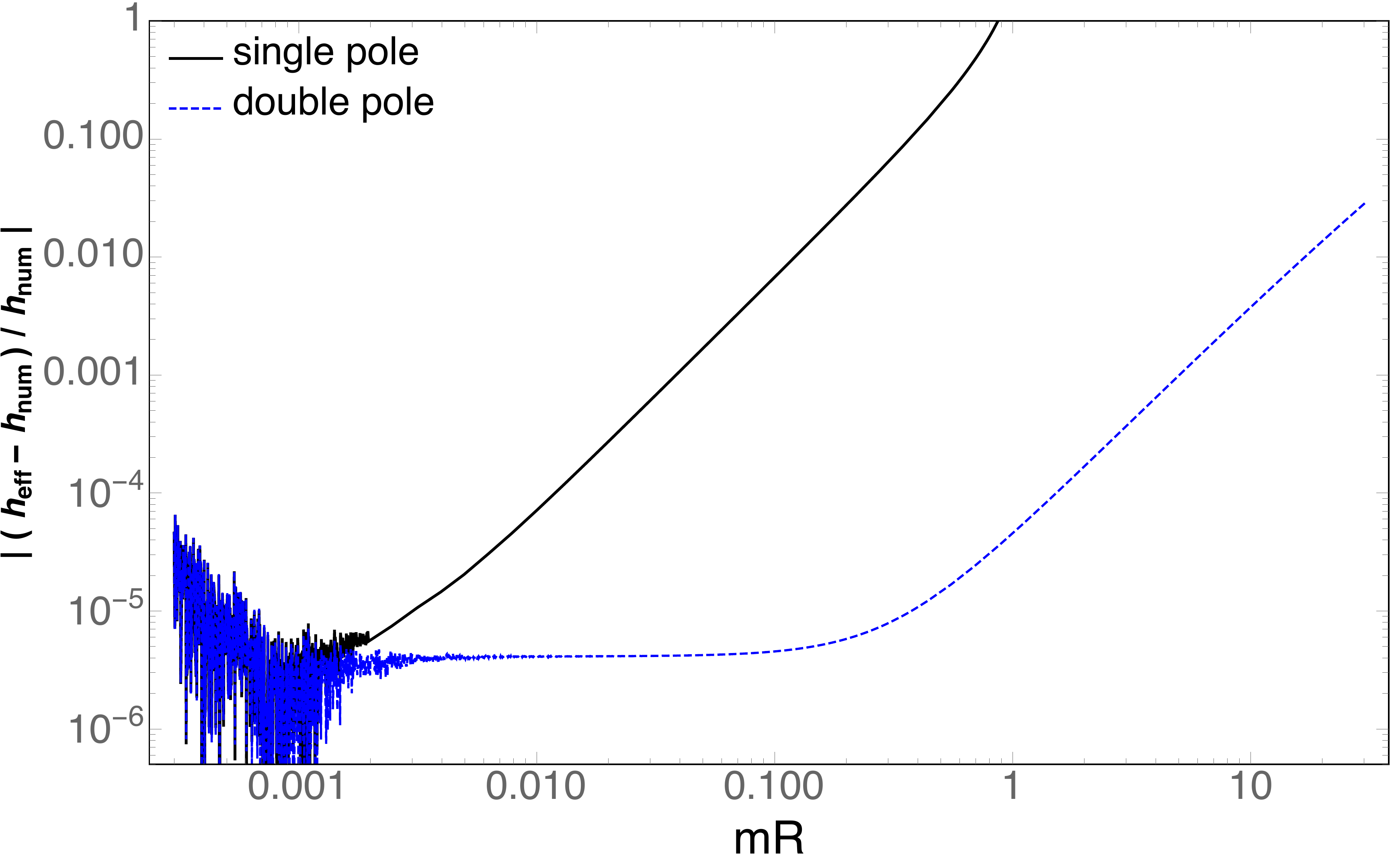} 
\caption{\small A plot of the relative error made when computing $\delta \omega/|\psi(0)|^2$ for nonzero $C/A$ using two analytic approximations single pole (black solid) and double-pole (blue dashed) to the right-hand side of eq.~\pref{ACGammastxt} as described in the main text. The plot's horizontal axis is $mR$, where $m$ is the mass of the orbiting fermion and $R$ is the size of the source. For plotting purposes we use $Z\alpha = 1/137$ and compute the shifts to the parity-even $j=\frac12$ state with $n=2$ assuming the source to be a shell of positive charge with radius $R$.}\label{Fig:multiplepole}
\end{center}
\end{figure}

The double-pole approximation turns out to be much better then the single-pole one (particularly given that the left-hand side, $\cQ := - (C/A)$, of \pref{ACGammastxt} turns out to be positive for small $\delta z$), and suffices for identifying the leading energy shift and its first subleading correction.  This can be seen in Fig.~\ref{Fig:multiplepole}, which compares the solution obtained for $\delta \omega$ using these approximate formulae to numerical results. For the purposes of these comparisons the source is assumed to be a fixed charged shell of radius $R$, whose energy eigenvalues can be computed exactly, and the state whose energy is perturbed is taken to be a parity-even $S$ state (similar results obtain for parity-odd states). The plots show that the error obtained when using the double-pole approximation is order $(Z\alpha)^2$ out to $mR \lsim \cO(1)$, for reasons identified below when we seek to compute $\cO(Z\alpha)^2$ terms. 

Concretely, the double-pole approximates the right-hand side of \pref{ACGammastxt} using the leading Laurent expansion near the poles of the Gamma functions,
\bea \label{newFdz2}
 G(x_\ssN+ \delta x)  &:=& \frac{\Gamma(1+2 \zeta)}{\Gamma(1-2\zeta)} \; \frac{\Gamma\left[ y(x)- 2\zeta\right]}{\Gamma\left[ y(x) \right]}  
 \simeq \frac{4(1-\zeta) \, \delta x}{(N+2)(N+1) (2 - 2\zeta -\delta x)} \,,
\eea
which uses $y(x) = \zeta - x$ and $x = Z\alpha\,\omega/\kappa$ and so  $y(x_\ssN + \delta x) = -N + \delta y = -N - \delta x$ where $x_\ssN = N +\zeta$ corresponds to the Dirac-Coulomb energy eigenvalue \pref{DiracE0txt} for a point source. To proceed we regard $\cQ := - (C/A)$ as a function of $f(\epsilon)/g(\epsilon)$ and $\omega$ and evaluate it at $\omega = \omega_{\ssN}$, equating the result to \pref{newFdz2}. This allows $\delta x$ (and hence also $\delta \omega$) to be solved for explicitly as
\be
 \delta x = \frac{  Z \alpha \,m^2 \delta \omega}{(m^2 - \omega_\ssN^2)^{3/2}}   \simeq   \frac{n (n+1)\cQ/2}{1 + n(n+1)\cQ/[2(Z\alpha)^2]} \,,
\ee
where (because our later focus is on $j = \frac12$) we trade $N$ for the principal quantum number, $n = N+1$, and write $1-\zeta \simeq \frac12 (Z\alpha)^2$. (The single-pole approximate differs from the above by taking the denominator to be unity, and only gives the leading contribution reliably in the limit $mR \ll 1$, if $R$ is the typical size of the source.)

It is useful to extract the naive Coulomb wave-function at the origin from $\delta \omega$ by writing
\be \label{heffdef}
  \delta \omega = \frac{h_{\rm eff}}\pi  \left(\frac{mZ\alpha}{n} \right)^3 \,,
\ee
where tracking through the definitions gives
\be \label{heffvsdx}
 h_{\rm eff}  = \frac{\pi c_n^{3/2} \delta x}{ Z \alpha \, m^2} \simeq  \frac{\pi c_n^{3/2} }{Z \alpha \, m^2} \left[ \frac{\frac12\,n (n+1)\cQ}{1 + \frac12\, n(n+1)\cQ/(Z\alpha)^2} \right] \,,
\ee
where we write $m^2 - \omega_{\ssN}^2 = c_n (Z\alpha m/ n)^2$ and so 
\be
 c_n^{3/2} = 1 + \frac{3 (n-1) (Z\alpha)^2}{2n^2}  + \cO(Z\alpha)^4 \,,
\ee
which can be taken as unity for the leading and $\cO(Z\alpha)$ correction but not once order $(Z\alpha)^2$ contributions are required. As we shall see, for $\cO(Z\alpha)^2$ corrections \pref{newFdz2} must also be revisited to include also subleading terms in $\delta x$.

\subsubsection*{Determining $\cQ = - C/A$}

To use the above formulae in practice we require an expression for how $\cQ = - C/A$ depends on the properties of the source. If the UV completion were a specific classical distribution, $\rho(r)$, of radius $R$ then $C/A$ would be fixed by demanding continuity of $f/g$ between the exterior and interior solutions at $r = R$ (examples of this are discussed in more detail below). In general, knowledge of $C/A$ is equivalent to knowledge of $f/g$ at some radius, since this is ultimately the only way the physics of the source influences exterior phenomena. 

What is required then is an explicit expression for $C_\pm/A_\pm$ as a function of $f_\pm(\epsilon)/g_\pm(\epsilon)$. In principle this is obtained by taking the ratio of expression \pref{fACgentxt} and \pref{gACgentxt} for the exterior solution (for each parity) and solving the resulting equations for $C_+/A_+$ and $C_-/A_-$. This is efficient and easy to implement numerically and once this is done $f_\pm/g_\pm$ at $r=\epsilon$ can be traded for constants in the source action through boundary conditions like \pref{cscvtofg}.

Analytic expressions\footnote{Such analytic expressions are useful (even when numerical results are easy) for tracking the leading parametric dependence of energy shifts on external variables.} for the required relation for $C_\pm/A_\pm$ can also be found when $\epsilon$ is small enough to justify keeping only the leading small-$r$ asymptotic form for the confluent hypergeometric functions in \pref{fACgentxt} and \pref{gACgentxt}. Specializing to states with $j= \frac12$ --- {\em i.e.} parity-even ($S$) states $f_+$ and $g_+$ and parity-odd ($P$) states $f_-$ and $g_-$ --- since these are the states most sensitive to finite-size effects of the source, we find the leading small-$\epsilon$ form
\bea \label{CAbc}
 \frac{C_\pm}{A_\pm}  &=& \frac{[\zeta \pm 1 + Z\alpha X]  - [(\zeta \mp 1)X - Z\alpha](f_\pm/g_\pm)}{[\zeta \mp 1 - Z\alpha X] - [(\zeta \pm 1)X + Z \alpha](f_\pm/g_\pm)} \; (2\kappa \epsilon)^{2\zeta} \,,
\eea
where $f_\pm/g_\pm$ is evaluated at $r = \epsilon$ and $X$ is defined by $X := \sqrt{{(m-\omega)}/{(m+\omega)}}$. As we shall see, it is the factor of 2 in the exponent of $(2\kappa \epsilon)^{2\zeta}$ that is responsible for the main differences between this Dirac case and the Klein-Gordon problem studied in \cite{PPEFT2} (for which instead ($2\kappa \epsilon)^{\zeta_s}$ appeared). This factor has its origins in the spin-orbit coupling that mixes two different orbital angular momenta into each state having fixed $j$. 

\subsection{Leading and first-subleading energy shifts}

For detailed studies of the influence of nuclei on atomic energy levels one expands all contributions to bound state energies as a dual series in the small parameters $(Z\alpha)^2$ and $m\epsilon Z\alpha \sim \epsilon/a_\ssB$, where $a_\ssB = 1/(mZ\alpha)$ is the Bohr radius and $\epsilon \simeq R$ where $R \simeq 1$ fm is a typical nuclear size. In practice, comparison with experiments on atomic energy levels requires both the leading contribution and its subleading $\cO(mRZ\alpha)$ correction, and for electronic atoms $(Z\alpha)^2$ corrections are also required since for $R$ of order a Fermi these are comparable in size to $(mRZ\alpha)$ corrections. Our purpose in this section is to identify as generally as possible how these terms depend parametrically on the properties of the source.

Although \pref{CAbc} is sufficient for some applications, a more accurate approximation turns out to be required in order to track the leading subdominant coefficients in this kind of expansion. Increased accuracy is required for bound-state calculations because nominally independent variables like $\kappa$ and $X$ become specific powers of $Z\alpha$ once evaluated at the lowest-order bound-state energies $\omega = \omega_{\ssN}$. For instance, using \pref{DiracE0txt} in the definitions implies
\be
 \rho_{nj} = 2 \kappa_{\ssN}\epsilon = \frac{2m\epsilon Z \alpha}{n} \Bigl[1 + \cO(Z\alpha)^2 \Bigr] \qquad \hbox{and} \qquad
 X_{nj} =  \frac{Z \alpha}{2n} \Bigl[1 + \cO(Z\alpha)^2 \Bigr] \,,
\ee
and so higher powers of these compete with powers of $Z\alpha$ arising elsewhere (such as from the expansion of $\zeta$). Extracting a particular order in $Z\alpha$ is further complicated by the appearance of factors of $(1-\zeta)^{-1} \propto (Z\alpha)^{-2}$ in the expansion of the confluent hypergeometric functions $\cM[a,1-2\zeta; \rho]$, due to the singularity of $\cM[a,b; z]$ as $b$ approaches a nonpositive integer.

We next identify the leading and subleading $\cO(mRZ\alpha)$ and $\cO(Z\alpha)^2$ contributions to the energy shift. To do so we use the exact expressions, \pref{fACgentxt} and \pref{gACgentxt}, for the general Dirac-Coulomb solution and solve for the integration constants $\cQ = - (C/A)$ in terms of $f/g$ evaluated at $r = \epsilon = R$, finding
\bea \label{Qform2}
 \cQ = - \; \frac{C}{A}  &=& \left\{ \left[ \frac{(Q_{20}+Q_{10})g +X(-Q_{20}+Q_{10})f }{(Q_{21}+Q_{11})g + X(-Q_{21}+Q_{11})f} \right] \rho^{2\zeta} \right\}_{r=R} 
\eea
where (as before) $X := \sqrt{(m-\omega)/(m+\omega)}$ and
\bea
  &&Q_{10} := \cM\left( \zeta - x,  1+2 \zeta; \rho \right) \,, \qquad
  Q_{11} :=  \cM\left( - \zeta - x , 1-2\zeta; \rho\right) \,,\nn\\
 &&\qquad\qquad Q_{20} := - \left( \frac{\zeta - x}{K - \hat x} \right) \cM\left( \zeta - x+1,  1+2 \zeta; \rho \right)  \\
 &&\hbox{and}\qquad\; Q_{21} := \left( \frac{\zeta + x}{K - \hat x} \right) \cM\left( - \zeta - x+1 , 1-2\zeta; \rho\right) \,,\nn
\eea
with $x = Z\alpha\, \omega/\kappa$ while $\hat x = Z \alpha \,m/\kappa$. These are to be evaluated at the lowest-order solution, $x = x_\ssN = N + \zeta$, where $N = n-1$ and $\zeta \simeq 1 + \frac12(Z\alpha)^2$ for $j=\frac12$ states, and we work only to subdominant order in $mRZ\alpha$ and $(Z\alpha)^2$. Eq.~\pref{Qform2} agrees with \pref{CAbc} at lowest order in $\rho$, for which $\cM[a,b; \rho] = 1$.

Since $\rho = 2\kappa_\ssN R \propto mRZ\alpha$ working to fixed order in $Z\alpha$ allows us to expand $\cM$ in powers of $\rho$, but when doing so must be careful about factors of $1/(1-\zeta) \propto (Z\alpha)^{-2}$ appearing in the coefficients of the hypergeometric series. Such terms only arise when $b$ of $\cM[a,b;\rho]$ is a negative integer and so only are a factor in $Q_{11}$ and $Q_{21}$. Since all powers of $\rho$ involve the factor $mR$ our guiding principle when expanding in $\rho$ is to keep terms involving only a single subdominant power of $Z\alpha$. This also allows us to neglect all subdominant powers of $1-\zeta \propto (Z\alpha)^2$ in any $\rho$-dependent terms. Using $\zeta \simeq 1 - \frac12 (Z\alpha)^2$ and $x \simeq x_\ssN = N+\zeta \simeq N+1$ one finds
\bea
  Q_{10} &:=& \cM\left( \zeta - x,  1+2 \zeta; \rho \right)  \simeq 1 - \left( \frac{N}{3} \right) \rho +  \frac{N(N-1)}{24} \; \rho^2 +\cdots  \,,
\eea
and
\bea
 Q_{20} &:=& - \left( \frac{\zeta - x}{K - x} \right) \cM\left( \zeta - x+1,  1+2 \zeta; \rho \right) \nn\\
 &\simeq& - \left( \frac{ N}{N+1-K} \right) \left[ 1 - \left( \frac{N -1}{3} \right) \rho +  \frac{(N-1)(N-2)}{24}\; \rho^2 + \cdots \right] \,,
\eea
while
\bea
  Q_{11} &:=&  \cM\left( - \zeta - x , 1-2\zeta; \rho\right) \nn\\
  &\simeq&  1 + \left( N+2 \right) \rho - \left[ \frac{(N+2)(N+1)}{2(1-\zeta)} \right] \frac{\rho^2}{2} +  \left[ \frac{N(N+2)(N+1)}{2(1-\zeta)} \right] \frac{\rho^3}{3!} + \cdots \,,
\eea
and
\bea
   Q_{21} &:=& \left( \frac{\zeta + x}{K - x} \right) \cM\left( - \zeta - x+1 , 1-2\zeta; \rho\right)\\
    &\simeq& -\left( \frac{ N+2}{N+1-K} \right) \left\{ 1 + \left( N+1 \right) \rho - \left[ \frac{(N+1)N}{2(1-\zeta)} \right] \frac{\rho^2}{2} \left[ \frac{N(N+1)(N-1)}{2(1-\zeta) } \right] \frac{\rho^3}{3!}  + \cdots \right\}\,.\nn
\eea

\subsubsection*{Parity-even leading energy shifts}

Collecting results and specializing to the parity-even $j=\frac12$ $S$ states ({\em i.e.} those with $K = -1$) gives the leading contribution (unsuppressed by any additional powers of $Z\alpha$)
\be \label{Qpllead}
  \frac12 \, n(n+1) \cQ_+ \simeq   \left[ \frac{2(1+2\xi_g)  }{1 - 2(1+2\xi_g )(mR)^2} \right] \left(  mRZ\alpha  \right)^2 \qquad \hbox{(leading order)} 
\ee
where $\xi_g$ contains the entire contribution of the physics of the source, through \eqref{xig}.

Using \pref{Qpllead} in the double-pole approximation \pref{heffvsdx} then gives
\bea \label{leadingeven}
 h_{\rm eff}^+   &\simeq&   \frac{\pi  }{Z \alpha \, m^2} \left[ \frac{\frac12 n (n+1)\cQ_+}{1 +\frac12 n(n+1)\cQ_+/(Z\alpha)^2} \right] \nn\\
 &=& 2\pi \, Z\alpha R^2 (1+2\hat g_1 )   \quad \hbox{(leading order)} \,,
\eea
where we use $\xi_g \simeq \hat g_1$ because at leading order consistency requires also dropping subleading terms in $\xi_g$.  Notice the cancellation here of the spurious $(mR)^2$ terms in the denominator of \pref{Qpllead}; a cancellation that is missed if only the single-pole approximation is used (thereby showing that physical energy shifts lie beyond its domain of validity).

\subsubsection*{Parity-even subleading $\cO(mRZ\alpha)$ energy shifts}

Including also subdominant terms linear in $Z\alpha$ requires keeping corrections coming from the expansion of the higher orders in $\rho$, leading to 
\be
  \frac12 \, n(n+1) \cQ_+ \simeq   \left[ \frac{1+2\xi_g  - \Delta_1^+ }{1 - 2(1+2\xi_g  - \Delta_1^+)(mR)^2+\Delta_2^+}  \right] 2\left(  mRZ\alpha  \right)^2 \qquad \hbox{(subleading order)}
\ee
where we use $\xi_g = \hat g_1 + \hat g_2 (mRZ\alpha)$ in the explicitly written terms, but it suffices to use only $\xi_g = \hat g_1$ in the quantities
\bea
  \Delta_1^+ &:=&  2(n-1) \left(\hat g_1 +    \frac{2n-1}{6n}\right) \frac{mRZ\alpha}{n}
  \nn\\ \hbox{and} \quad
  \Delta_2^+ &:=& \Bigl[ 1 + 2n(1+\hat g_1 ) \Bigr] \frac{mRZ\alpha}{n} \,.
\eea
Consequently the double-pole approximation gives
\bea \label{subleadingeven}
 h_{\rm eff}^+   &\simeq&   \frac{\pi  }{Z \alpha \, m^2} \left[ \frac{\frac12 n (n+1)\cQ_+}{1 +\frac12 n(n+1)\cQ_+/(Z\alpha)^2} \right] \nn\\
 &\simeq& 2\pi \, Z\alpha R^2 \Bigl[ (1+2\hat g_1)(1-\Delta_2^+) + 2\hat g_2(mRZ\alpha) - \Delta_1^+  \Bigr] \\
  &=& 2\pi \, Z\alpha R^2 \left\{ 1+2\hat g_1 + 2\hat g_2(mRZ\alpha) - \left[ 1+8n^2 \left( 1 + \frac32\,\hat g_1(\hat g_1+2) \right) \right] \frac{mRZ\alpha}{3n^2} \right\}   \,,  \nn
\eea
which includes all corrections that are down only by a single power of $Z\alpha$ (but drops $(Z\alpha)^2$ everywhere). Later sections verify that these expression capture specific special cases in the literature.

\subsubsection*{Parity-odd leading energy shift}

We next turn to parity-odd $j=\frac12$ $P$ states (for which $K = +1$). In this case following the same steps reveals the leading contribution to be
\be
  \cQ_- \simeq -\left( \frac{n-1}{2n} \right) \left[ \xi_f  - \frac23 (mRZ\alpha)  \right] \left( \frac{2mRZ\alpha}{n} \right)^2 \qquad \hbox{(leading order)} 
\ee
where the entire contribution of source physics is through 
\be
 X \left( \frac{f_-}{g_-} \right)_{r=R} = \frac{\xi_f}{2n} \qquad \hbox{with} \qquad
 \xi_f = \hat f_1 (mRZ\alpha) + \hat f_2(mRZ\alpha)^2+ \hat f_3(Z\alpha)^2 + \cdots \,.
\ee
with (as before) $X = \sqrt{(m-\omega)/(m+\omega)}$. 
 
Dropping all subdominant powers of $Z\alpha$ (and for consistency restricting the source contribution to $\xi_f \simeq \hat f_1 (mRZ\alpha)$ gives the leading parity-odd energy shift
\bea
 h_{\rm eff}^-  & \simeq&    -\; \frac{\pi(n^2-1)}{n^2} \left( \hat f_2  - \frac23 \right)  (Z\alpha)^2 m R^3 \qquad \hbox{(leading order)}  \,.
\eea
As usual, this is smaller than the parity-even result because it is suppressed by the spin-orbit coupling required to link the $P$ states to $\ell=0$ orbital angular momentum.

\subsubsection*{Parity-odd subleading $\cO(mRZ\alpha)$ energy shift}

Even though small, for some special cases (such as the charged shell described below) it happens that $\hat f_1 = \frac23$ and so the leading contribution to parity-odd states vanishes. Such cases are dominated by the subleading contribution, for which
\be
  \frac12 \, n(n+1) \cQ_- \simeq -\; \frac{n^2-1}{n^2}  \left[ \frac{\xi_f  - \frac23(mRZ\alpha) + \Delta_1^- }{1 + (n^2-1) (mR/n)^2 [\xi_f  - \frac23(mRZ\alpha)]- \Delta_2^-} \right] \left(  mR Z\alpha  \right)^2 
\ee
where we can use $\xi_f = \hat f_1 (mRZ\alpha) + \hat f_2 (mRZ\alpha)^2$ in the explicitly written factors, but stop at $\xi_f \simeq \hat f_1(mRZ\alpha)$ in
\bea
  \Delta_1^-  
  &\simeq& \Bigl[  (n-2)  - (2n-3)\hat f_1  \Bigr] \frac{ (mRZ\alpha)^2}{3n} \\
  \hbox{and} \qquad \Delta_2^- &=& 
   \frac{1}{2} \left[ \hat f_1  -  \frac{2(n+1)}{n} \right] (mRZ\alpha) \,,\nn
\eea
leading to
\bea
 h_{\rm eff}^-   &\simeq&   \frac{\pi  }{Z \alpha \, m^2} \left[ \frac{\frac12 n (n+1)\cQ_-}{1 +\frac12 n(n+1)\cQ_-/(Z\alpha)^2} \right] \nn\\
 &\simeq&  -\; \frac{\pi(n^2-1)}{n^2}  Z\alpha R^2 \left[ \frac{ \xi_f  - \frac23(mRZ\alpha) + \Delta_1^- }{ 1 -\Delta_2^-  } \right] \quad\hbox{(subleading order)} \\
&\simeq&  -\; \frac{\pi(n^2-1)}{n^2}  (Z\alpha )^2 m R^3 \left\{  \left( \hat f_1  - \frac23\right) \left[ 1 + \left( \frac{\hat f_1}{2}  -\frac{5}{3}\right) (mRZ\alpha) \right]   +\left( \hat f_2 -\frac{1}{9} \right) (mRZ\alpha)    \right\}  \,.\nn
\eea

\subsection{Subleading $( Z\alpha)^2$ energy shifts}

This section computes the subdominant $\cO(Z\alpha)^2$ energy shifts for parity even and parity odd cases. Because factors of $mR$ do not accompany the subleading powers of $Z\alpha$ it suffices to drop all nontrivial powers of $\rho$ from the get-go and instead focus on the subdominant powers of $(Z\alpha)^2$. Because of this we can evaluate $\cQ$ directly using \pref{CAbc}, which is repeated here for convenience
\bea
 \cQ &\simeq& 
 \frac{[K-\zeta - Z\alpha X]g +[(K+\zeta)X - Z\alpha]f }{[K+\zeta- Z\alpha X]g + [(K-\zeta)X- Z\alpha ]f} \; (2\kappa R)^{2\zeta} \,.
\eea
This is to be expanded to order $(Z\alpha)^2$, using $\zeta \simeq 1 - \frac12(Z\alpha)^2$ and
\bea
 \omega \to \omega_\ssN &=& m \left[ 1 + \frac{(Z \alpha)^2}{ \left(N + \zeta \right)^2} \right]^{-1/2}
  \simeq m \left[ 1 - \frac{(Z\alpha)^2}{2n^2} - \frac{(4n-3)(Z\alpha)^4}{8n^4} +\cO[(Z\alpha)^6 \right]  \,,
\eea
and
\bea
 \kappa \to \kappa_\ssN &=& \sqrt{(m-\omega_\ssN)(m+\omega_\ssN)} 
 \simeq m\left[ \frac{Z\alpha}{n} + \frac{(n-1)(Z\alpha)^3}{2n^3} + \cO[(Z\alpha)^5] \right]\,, 
\eea
so in particular
\be 
 (2\kappa_\ssN R)^2 \simeq \left[ 1+ \frac{(n-1)(Z\alpha)^2}{n^2} + \cO[(Z\alpha)^4] \right] \left(\frac{2mRZ\alpha}{n} \right)^2\,.
\ee
Similarly
\be
 X \to X_\ssN =  \sqrt{\frac{m-\omega_\ssN}{m+\omega_\ssN}} 
 \simeq \frac{Z\alpha}{2n} + \frac{(2n-1)(Z\alpha)^3}{8n^3} + \cO[(Z\alpha)^5] \,,
\ee
Using the corresponding terms in the source expansion $\xi_g \simeq \hat g_1 + \hat g_3 (Z\alpha)^2$ then gives $\cQ_+$ for parity-even ($K=-1$) states as
\bea
  \frac{n(n+1)\cQ_+}{2}      &\simeq&  \left(mR Z\alpha\right)^2\left\{ 2(1+2\hat g_1)\left[1- (Z\alpha)^2   \ln \left( \frac{2mRZ\alpha}{n} \right) \right]   \right. \nn\\
  && + \left[ \frac{(6n^2-n-3) -(2n^3-4n^2+n+3)2\hat g_1  -4n^2(n+1)\hat g_1^2}{2n^2(n+1)} \right](Z\alpha)^2 \nn\\
  && \qquad\qquad\qquad\qquad\qquad \left. \phantom{\frac12}+ 4\hat g_3(Z\alpha)^2+ \cO[(Z\alpha)^4]\right\} \,.
\eea

\begin{table}[htp]
\caption{First few harmonic numbers}
\begin{center}
\begin{tabular}{c|ccccccccc}
$N$ &0&1 &2&3&4&5&6&7&8 \cr
\hline\hline
$H_\ssN$ &0& 1& $3/2$ & $11/6$ & $25/12$&$137/60$&$49/20$&$363/140$&$761/280$
\end{tabular}
\end{center}
\label{Tab:harmonic}
\end{table}%

To work systematically to relative order $(Z\alpha)^2$ we must keep track of the factor of $c_\ssN$ in $h_{\rm eff}$ 
\be
 h_{\rm eff}   \simeq \frac{\pi c_\ssN^{3/2} \delta x}{ Z \alpha \, m^2} \simeq  \frac{\pi \delta x }{ Z \alpha \, m^2} \left[1+ \frac{3(n-1)(Z\alpha)^2}{2n^2} \right] \,,
\ee
and it is also necessary to refine the double-pole approximation, by keeping subdominant terms in the Gamma-function expansion:
\be
   \Gamma(y) = \Gamma(\delta y - N) \simeq \frac{(-)^N}{N!} \left[ \frac{1}{\delta y} + H_\ssN - \gamma  + \cO(\delta y) \right] \,,
\ee
where the harmonic numbers (see also Table \ref{Tab:harmonic}) are defined by
\be
 H_\ssN = \sum_{k=1}^N \frac{1}{k} = \int_0^1 \exd x \; \frac{1-x^N}{1-x} \,,
\ee
and the integral representation shows in particular that $H_0 = 0$. $\gamma$ is the Euler-Mascheroni constant 
\be
 \gamma = \lim_{N\to \infty} \Bigl[ H_\ssN - \ln N \Bigr] = 0.57721\,56649\,01532\,86060\,65120\,... \nn
\ee

Tracking only the $m$-independent $(Z\alpha)^2$ terms the leading contributions then are
\bea
 h_{\rm eff}^+ &\simeq&  \pi Z\alpha \, R^2 \left\{ 2(1+2\hat g_1) \left[ 1 - (Z\alpha)^2 \left[   \ln \left( \frac{2mRZ\alpha}{n} \right)+ H_{n+1} +\gamma \right]\right]\right.\\
  &&\qquad\qquad \left.+ \left[ 4\hat g_3+5 +8\hat g_1  -2\hat g_1^2  + (1+2\hat g_1)\frac{12n^2-n-9}{2n^2(n+1)} \right](Z\alpha)^2  + \cO[(Z\alpha)^4]\right\} \,. \nn
\eea
The first term agrees with the leading result found earlier, and to these can be added the subleading $(mRZ\alpha)$ corrections found in eq.~\pref{subleadingeven} above.

Some implications of these formulae are explored in the next sections.

\section{Examples} \label{examples_sec}

As ever, the power in using an effective action to describe the short-distance properties of the source lies in its generality. That is, coefficients like $c_s$, $r_p$ and $c_v$ can be used to describe the leading contributions due to {\em any} localized source physics, provided only that this physics arises over small enough scales, $R$, to make an expansion in powers of $R/a$ useful (where $a$ is a typical macroscopic scale --- such as the Bohr radius of an exterior orbit). This ensures the model-independence of parametrizing physical quantities like energy shifts in terms of these parameters. 

This section emphasizes this point by indicating how several kinds of microscopic source physics contribute to effective couplings in the source action, $S_p$, and how the above expressions reproduce familiar results in specific instances.

\subsection{Explicit charge distributions}

Perhaps the simplest example of microscopic source physics that can be parametrized by $S_p$ is the situation where the source is an explicit static charge distribution, $\rho(x)$, rather than a point charge. Examples of this form are studied in the literature, with sensitivity to source structure often estimated by tracking how energy shifts alter as $\rho(x)$ is varied through a plausible range of configurations \cite{Muonic, Eides, Friar, Nickel, NASA}. 

\subsubsection{Relations to moments}

The leading terms in the source-dependent energy shift in this case have been calculated by perturbing the interior solution around the Coulomb problem and are known\footnote{See also \cite{Franziska} for a discussion of the limits of this expansion.} to be given by \cite{Friar}
\be \label{hvsmoments}
 h_{\rm eff} = \frac{2\pi}3 \, Z\alpha \left[ r_p^2 - \frac{Z\alpha \mu}{2} \langle r^3 \rangle_{(2)} + (Z\alpha \mu)^2 F_{\scriptscriptstyle NR} + (Z\alpha)^2 F_{\scriptscriptstyle REL} \right] \,,
\ee
where $\mu$ is the reduced mass (so $\mu \to m$ in the infinite-source-mass limit used here) and
\be
 \langle r^3 \rangle_{(2)} = \int \exd^3 x \, \exd^3y \, |\bfx|^3 \rho(\bfy-\bfx) \rho(\bfy) \,,
\ee
and $F_{\scriptscriptstyle NR}$ and $F_{\scriptscriptstyle REL}$ are given in terms of various charge moments in \cite{Friar}. 

This result is model-independent inasmuch as the expression for the coefficients of the series are universal functions of these moments, and so with the energy shift due to various charge distributions just differing in the values these distributions predict for the moments themselves. This is a more limited sense of `model-independence' than we use here, since the model-independence of the predictions of the effective action apply not just to static charge distributions, but essentially to {\em any} kind of source physics that is sufficiently localized. (This model-independence of EFT methods for atomic measurements are emphasized within the 2nd-quantized framework in \cite{HillPaz, Contact}.) 

We verify in Appendix \ref{App:DiracSolutions} that for a general static charge distribution, $\rho(x)$, the quantity $\hat g_1$ that dominates how source physics appears in $g_+/f_+$ is related to the rms charge density, $r_p^2 = \langle r^2 \rangle$, by
\be \label{rpvsg1}
 (1+2\hat g_1)R^2 = \frac{r_p^2}3 \,,
\ee
which implies that the leading energy shift given by \pref{leadingeven} becomes
\be
 h_{\rm eff}^+ \simeq \frac{2\pi}3 \, Z\alpha \, r_p^2 \,,
\ee
as required for consistency with \pref{hvsmoments}. On the other hand, the boundary condition \pref{cscvtofg} shows how the parameter $\hat g_1$ is also interchangeable with one combination of $c_s$ and $c_{v\,{\rm tot}}$ through
\be \label{csvsg1}
 \Big( c_s + c_{v\,{\rm tot}} \Bigr)_{\epsilon = R} = c_s + c_v + \frac{2\pi}3 \, Z\alpha \, r_p^2 = 4\pi R^2 \left( \frac{g_+}{f_+} \right)_{r = R} = 4\pi \hat g_1 \, Z\alpha \, R^2\,.
\ee
This implies 
\begin{equation}
 c_s + c_v = - 2 \pi Z\alpha\, R^2\,,
\end{equation}
{\em i.e.} the infrared fixed point found in \eqref{cs+cvIR}. Note the difference from the Schr\"odinger running where we found that $h= 0$ is a fixed point that parametrizes a trivial boundary condition.

The subdominant $(mRZ\alpha)$ contribution also provides a relation between $\hat g_2$ and the higher moment $\langle r^3 \rangle_{(2)}$. Comparing \pref{subleadingeven} with \pref{hvsmoments} and using \pref{rpvsg1} shows 
\bea \label{subleadingevenex}
 \langle r^3 \rangle_{(2)}  \simeq -6 R^3 \left\{  2\hat g_2 - \left[ 1+8n^2 \left( 1 + \frac32\,\hat g_1(\hat g_1+2) \right) \right] \frac{1}{3n^2} \right\}   \,, 
\eea
Although we do not have a general proof of this result, we can verify it for specific charge distributions. These higher terms can be related to higher-dimension interactions --- such as those of \pref{nexdim} --- in $S_p$, using matching conditions similar to \pref{csvsg1}, although we do not pursue this here.

\subsubsection{Specific charge distributions}

The detailed calculations done for specific charge distributions \cite{Friar, Nickel, NASA} provide useful checks on the higher-order terms, since these must agree on the series coefficients for the specific charge distributions studied. To provide this check we compute the couplings $\hat g_i$ and $\hat f_i$ for various charge distributions in Appendix \ref{App:DiracSolutions}, and we here use these in the above expressions for $h_{\rm eff}$ to verify agreement where overlap is possible.  

\subsubsection*{Spherical charged shell}

The simplest such example is that of a charged shell, for which
\be
 \rho = \sigma \, \delta(r - R) = \frac{Ze}{4\pi R^2} \, \delta (r-R)\,
\ee
which is convenient since the interior solution can be solved exactly in closed form. (We have checked that our numerical results for this case agree with those of \cite{NASA}.) For this distribution the rms charge radius is $r_p^2 = R^2$ and $\langle r^3 \rangle_{(2)} = 16R^3/5$. 

For the parity-even state the boundary parameters appearing in $g_+(R)/f_+(R)$ work out to be
\be
  \hat g_1 = - \, \frac13 \,, \qquad \hat g_2 = - \frac{2}{45} + \frac{1}{6n^2}  \qquad \hbox{and} \qquad \hat g_3 = - \,\frac{1}{45} \,,
\ee
while for the parity-odd state the analogous parameters are
\be
  \hat f_1 = + \frac23 \,, \qquad \hat f_2 = +  \frac{2}{45} \qquad \hbox{and} \qquad \hat f_3 = + \frac{1}{3}  \,.
\ee

Using these values to compute the leading and subleading $(mRZ\alpha)$ and $(Z\alpha)^2$ energy shifts then gives 
\bea
  h_{\rm eff}^+  &\simeq&  \pi Z\alpha \, R^2 \left\{ 2(1+2\hat g_1) \left[ 1 - (Z\alpha)^2 \left[   \ln \left( \frac{2mRZ\alpha}{n} \right)+ H_{n+1} +\gamma \right]\right]\right.\nn\\
  &&\quad + 4\hat g_2(mRZ\alpha) - \left[ 1+8n^2 \left( 1 + \frac32\,\hat g_1(\hat g_1+2) \right) \right] \frac{2mRZ\alpha}{3n^2} \\
  &&\quad\quad \left.+ \left[ 4\hat g_3+5 +8\hat g_1  -2\hat g_1^2  + (1+2\hat g_1)\frac{12n^2-n-9}{2n^2(n+1)} \right](Z\alpha)^2  + \cO[(mRZ\alpha)^2, \, mR(Z\alpha)^2, \, (Z\alpha)^4]\right\} \nn\\
&\to&\frac{2\pi Z\alpha R^2}{3} \left\{1- \frac85  \Bigl( mRZ\alpha \Bigr) - \left[\ln \left(\frac{2m R\,Z\alpha}{n} \right)+ H_{n+1} + \gamma - \frac{91}{30} - \frac{12n^2-n-9}{4n^2(n+1)}  \right] \Bigl(Z\alpha \Bigr)^2 + \cdots \right\} \nn\\
&&\qquad\qquad \hbox{(charged shell)}\,,\nn
\eea
for parity-even states. Notice the correct result for $r_p^2$ and the cancellation of the $n$-dependence (and agreement with) the second moment $\langle r^3 \rangle_{(2)}$ for this distribution. This expression also agrees well with numerical evaluation (as illustrated in Fig.~\ref{Fig:multiplepole}). 

In this case, because $\hat f_1 = \frac23$, the leading parity-odd energy shift vanishes, leaving a result that is smaller than would naively be expected. The energy shifts predicted by the parameters $\hat f_i$ in this case are
\bea
 h_{\rm eff}^-  &\simeq&  -\; \frac{\pi(n^2-1)}{n^2}  (Z\alpha )^2 m R^3 \left\{  \left( \hat f_1  - \frac23\right) \left[ 1 + \left( \frac{\hat f_1}{2}  -\frac{5}{3}\right) (mRZ\alpha) \right]   +\left( \hat f_2 -\frac{1}{9} \right) (mRZ\alpha)    \right\}  \nn\\
&\to&  +\; \frac{\pi(n^2-1)}{45n^2}  (Z\alpha )^3 m^2 R^4   \qquad\qquad\hbox{(charged shell)}\,.
\eea
Both of these results also depend on $n$ in the way indicated by numerical evaluation.

\subsubsection*{Uniform spherical distribution}

A second go-to example is the case of uniform charge distribution, although in this case the interior solution cannot be computed in closed form. We have verified that our solutions agree in this case with the numerical results given in \cite{NASA}. Analytic expressions for the series expansion for the energy shifts are also given in \cite{Nickel}, and we have verified that our results agree with these (and with \cite{Friar}) in this case.

Evaluating the boundary condition $g_+(R)/f_+(R)$ using the interior solutions returns the following values
\be
  \hat g_1 = - \, \frac25 \,, \qquad \hat g_2 =- \frac{116}{1575}  + \frac{1}{6n^2} \qquad \hbox{and} \qquad \hat g_3 = - \frac{736}{17325}  \,,
\ee
while the same calculation for the parity-odd states gives
\be
  \hat f_1 =  \frac23 \,, \qquad \hat f_2 = +\frac{32}{315}\qquad \hbox{and} \qquad \hat f_3 = + \frac{2}{5}  \,.
\ee
Used in the parity-even energy shift, these values return the leading and sub-leading results
\bea
  h_{\rm eff}^+  &\simeq&  \pi Z\alpha \, R^2 \left\{ 2(1+2\hat g_1) \left[ 1 - (Z\alpha)^2 \left[   \ln \left( \frac{2mRZ\alpha}{n} \right)+ H_{n+1} +\gamma \right]\right]\right.\nn\\
  &&\quad + 4\hat g_2(mRZ\alpha) - \left[ 1+8n^2 \left( 1 + \frac32\,\hat g_1(\hat g_1+2) \right) \right] \frac{2mRZ\alpha}{3n^2} \\
  &&\quad\quad \left.+ \left[ 4\hat g_3+5 +8\hat g_1  -2\hat g_1^2  + (1+2\hat g_1)\frac{12n^2-n-9}{2n^2(n+1)} \right](Z\alpha)^2  + \cO[(mRZ\alpha)^2, \, mR(Z\alpha)^2, \, (Z\alpha)^4]\right\} \nn\\
&\to&\frac{2\pi Z\alpha R^2}{5} \left\{ 1- \frac{80}{63}  \Bigl( mRZ\alpha \Bigr) - \left[\ln \left(\frac{2m R\,Z\alpha}{n} \right)+ H_{n+1} + \gamma - \frac{22697}{6930} - \frac{12n^2-n-9}{4n^2(n+1)}  \right] \Bigl(Z\alpha \Bigr)^2 + \cdots \right\} \nn\\
&&\qquad\qquad \hbox{(uniform sphere)}\,.\nn
\eea
These agree with the coefficients given explicitly in \cite{Nickel}. The first two terms also agree with \cite{Friar} since the rms radius is $r_p^2 = \frac35 \, R^2$ for this distribution, while the second moment is $\langle r^3 \rangle_{(2)} = \frac{32}{21} \, R^3$ and so 
\be
 \frac{Z\alpha m \langle r^3 \rangle_{(2)}}{2r_p^2} = \frac12\; \frac{32}{21} \; \frac53 \; (mRZ\alpha) = \frac{80}{63}(mRZ\alpha) \,.
\ee

\subsection{Other applications}

A point-particle effective action like $S_p$ can be used to parametrize {\em any} short-range source physics and so need not be limited to describing the effects of finite nuclear size. This section summarizes a few such examples.

\subsubsection*{Vacuum polarization}

A standard contribution to atomic energy levels that also can be captured using $S_p$ is the contribution (or parts of the contribution) due to vacuum polarization.  It is well-known that the effects of vacuum polarization on the field of a point charge, $Ze$, can be described by the Ueling potential \cite{Ueling}, of the form
\be
 U(r) = \frac{2\alpha\, Ze}{3\pi r} \int_1^\infty \frac{\exd u}{u^2} \sqrt{u^2 -1} \left( 1 + \frac{1}{2u^2} \right) e^{-2ur/\alpha} \,,
\ee
in which $m$ is the mass of the particle circulating within the loop. Since the range of this interaction is of order $R \sim m^{-1}$ the electron and muon vacuum polarizations fall into the category of physical effects acting over much smaller distances than typical sizes of orbits in ordinary atoms. The same is true for the influence of the muonic vacuum polarization within muonic atoms (but because $m_e \sim \alpha m_\mu$ it is not true for the shifts on muonic atom energies due to electron vacuum polarization).

Such a potential shifts the energy of atomic states with low angular momentum that sample the potential near the nucleus, by an amount that is proportional (in the Schr\"odinger limit) to the wave-function at the origin: $|\varphi(0)|^2$. Using the notation of earlier sections, the resulting energy shift has size 
\be 
  h_{\rm eff} = - \, \frac{4Z\alpha^2}{15 m^2} \,,
\ee
where $m$ is the mass of the particle in the loop. Since the photon line of the vacuum polarization does not flip helicity the arguments of earlier sections imply that this leading energy shift is correctly captured (at order $(Z\alpha)^2/m^2$) in all low-energy observables through a contribution to the effective couplings in \pref{sourceaction} of size 
\be
  c_s = 0 \qquad \hbox{and} \qquad c_{v\,{\rm tot}} =  - \, \frac{4Z\alpha^2}{15 m^2} \,. 
\ee

\subsubsection*{Strong interactions and anti-protonic atoms}

When the particle orbiting a nucleus experiences the strong interaction (such as for a $\pi^-$, $K^-$ or $\bar p$) then it experiences a short-range ($R \sim m_\pi^{-1}$) strong interaction with the nucleus in addition to the usual Coulomb interaction. These are often described in the literature in terms of explicit nuclear potentials, which though concrete introduce an element of model-dependence into the treatment. 

For such situations a more model-independent approach is to use the contact interactions appearing in \pref{sourceaction} to capture the effects of these strong interactions on energy shifts and nuclear scattering amplitudes. This has the advantage of using only the short range of the force to organize the calculation, and so allows the disentangling of effects that rely only on this from those that instead depend on the detailed form assumed for any hypothetical nuclear potential. 

Ref.~\cite{PPEFT2} shows how parametrizing these strong interactions in terms of the lowest-dimension contact interaction allows the derivation of a relation between the strong-interaction induced shifts in atomic energy levels and the scattering length  for collisions with the nucleus, that reproduce the standard Deser formula \cite{Deser} (derived using nuclear potentials in the 1950s). 

The leading effects of the nuclear force on antiprotons in protonium \cite{Antiproton, AntiprotonExp} can similarly be captured through the contact interactions of \pref{sourceaction}, though for protonium the existence of a relatively quick annihilation channel reduces the practical utility of using measurements of the energy shifts to learn about the nuclear interaction. But because this annihilation can also be described in the effective point-source action through the addition of imaginary parts to the effective couplings $c_s$ and $c_v$ one use for $S_p$ in this case is to compute the dependence of the annihilation rate on the principal quantum number $n$ for $S$ and $P$ states. Thinking of the annihilation rate as the imaginary part of the energy eigenvalue shows that this $n$-dependence should be the same as for the energy shifts found in earlier sections, and this indeed reproduces what is found when modelling annihilation using nuclear potentials \cite{AntiprotonAnn}. 

The virtue of rederiving this result using $S_p$ is that the effective field theory shows why the result is robust, and not an artefact of model-dependent details. 

\subsubsection*{Exotic interactions}

A fairly obvious use for contact interactions in the point-particle action is to parametrize the effects of any hypothetical new forces acting between nuclei and electrons or muons, and in particular forces that differ in strength between these two (since these can be captured through species-dependent values for $c_s$ and $c_v$, unlike for $r_p$). Indeed the observation that the existence of such short-range interactions could, in principle, explain the proton radius puzzle \cite{Puzzle} has led to efforts to better understand their size \cite{Contact} and to the proposal of exotic interactions of this type \cite{Exotica}. 

\section{Summary} \label{sum_sec}

In this paper, we introduce the PPEFT of Dirac fermions using a first-quantized language for the heavy compact object and a second-quantized language for the lighter fermion with which it interacts. This formalism can be advantageous to the fully second-quantized framework in the limit of the compact object being much heavier than the light interacting particle, i.e.~the heavy compact object can be regarded as being in a position eigenstate to first approximation. 

This formalism was previously introduced for bosons \cite{PPEFT1,PPEFT2} where it was found that energy shifts due to the finite size $R$ of the source scale linearly in $R$ which is unusual. This does not carry over to fermions, i.e.~energy shifts scale as $R^2$. The absence of such unusual energy shifts means that there is no additional term that could account for the proton-radius-puzzle.

Our PPEFT allows one to parametrize the currently measurable energy shifts and their leading corrections due to the finite size of the nucleus for hydrogen and muonic hydrogen. Other applications include parametrizing strong interactions between the orbiting particle and the nucleus and anti-protonic atoms as well as hypothetical new forces acting between nuclei and electrons or muons. 

In general, energy shifts are found by comparing the ratio of integration constants, $C/A$, appearing in the mode expansions \pref{fACgentxt} and \pref{gACgentxt} for the radial solutions to the Dirac equation found in two ways. On one hand normalizability at large $r$ implies $C/A$ is given by \pref{ACGammastxt}, while on the other hand it is fixed by the boundary condition for the ratio of radial functions, $f/g$, evaluated at a small radius $r = \epsilon$ near where the small source intervenes. The expression for $C/A$ given $\left. f/g\right|_{r=\epsilon}$ can be found either by working numerically with the exact mode functions, or analytically using \pref{CAbc} if $\epsilon$ is small enough that the mode functions are well-approximated by their small-$r$ asymptotic forms. 

The main contribution of the PPEFT construction given here is to express $f/g$ at $r = \epsilon$ in terms of general effective couplings, such as $c_s$ and $c_v$ using the conditions given in eqs.~\pref{cscvtofg}. This leads to a low-energy expansion applicable to a {\em generic} source physics provided only that the size of the source is sufficiently small.  In the explicit calculations presented here `generic' is in practice restricted for simplicity to parity conserving and rotationally invariant sources, rather than considering different source models one at a time. Results for specific models of the source can then be found by evaluating $c_s$ and $c_v$ explicitly using the model, such as along the lines as was done in the text for specific charge distributions.

What sets the size of $\epsilon$? The above procedure works for boundary conditions provided at {\em any} small radius $r = \epsilon$, provided that $\epsilon$ is much smaller than the applications of interest (such as the Bohr radius, for atomic examples) while also being larger than the actual size $R$ of the source. The effective couplings --- {\em e.g.} $c_s$ and $c_v$ --- themselves also depend on $\epsilon$ in precisely the way required to ensure that physical quantities do not; an evolution computed for $c_s$ and $c_v$ explicitly in \S\ref{sec:RG}. Once $c_s$ and $c_v$ are specified by matching to a specific model at $r =R$, their size at larger $r=\epsilon$ is dictated by this evolution.

Finally, we give explicit formulae for energy shifts in the Dirac-Coulomb case as a double series in powers of $mRZ\alpha$ and $(Z\alpha)^2$, given a similar expansion for the boundary conditions $f/g$ of the form
\begin{equation}
		\frac{1}{Z\alpha}\left( \frac{g_+}{f_+} \right)_{r=R} = \hat g_1 + \hat g_2(mRZ\alpha) + \hat g_3 (Z\alpha)^2 + \dots\,, \ee
and
\be
  2n\sqrt\frac{m - \omega}{m + \omega}\left( \frac{f_-}{g_-} \right)_{r=R} = \hat f_1(mRZ\alpha) + \hat f_2 (mRZ\alpha)^2 + \hat f_3 (Z\alpha)^2 + \dots \,,
\end{equation}
with `plus' and `minus' referring to positive and negative parity eigenstates. The parameters $\hat f_i$ and $\hat g_i$ can be determined directly from a particular model of the underlying source and can be traded for parameters in the effective Lagrangian parameters (like $c_s$ and $c_v$, with higher orders also depending on their higher-dimensional counterparts).

Given such a boundary condition we write the energy shift to electrostatic bound states in terms of an effective $\delta$-function potential:
\be 
  \delta \omega^\pm = \frac{h^\pm_{\rm eff}}\pi  \left(\frac{mZ\alpha}{n} \right)^3 \,, 
\ee
where the effective coupling $h^\pm_{\text{eff}}$ is given order by order in $(Z\alpha)^2$ and $(mRZ\alpha)$ by:
\bea
 h_{\rm eff}^+ &\simeq&  \pi Z\alpha \, R^2 \left\{ 2(1+2\hat g_1) \left[ 1 - (Z\alpha)^2 \left[   \ln \left( \frac{2mRZ\alpha}{n} \right)+ H_{n+1} +\gamma \right]\right]\right. \nn\\
  &&\qquad\qquad \left.+ \left[ 4\hat g_3+5 +8\hat g_1  -2\hat g_1^2  + (1+2\hat g_1)\frac{12n^2-n-9}{2n^2(n+1)} \right](Z\alpha)^2\right.\\
  &&\qquad\qquad \left.+\left[ 2\hat g_2 - \frac{1}{3n^2} \left[ 1+8n^2 \left( 1 + \frac32\,\hat g_1(\hat g_1+2) \right) \right]\right](mRZ\alpha) + \dots \right\}\,, \nn
\eea
and
\bea
 h_{\rm eff}^-   &\simeq&  -\; \frac{\pi(n^2-1)}{n^2}  (Z\alpha )^2 m R^3 \left\{  \left( \hat f_1  - \frac23\right) \left[ 1 + \left( \frac{\hat f_1}{2}  -\frac{5}{3}\right) (mRZ\alpha) \right]\right. \nn\\
			&&\qquad\qquad\qquad\qquad\qquad\quad\left.+\left( \hat f_2 -\frac{1}{9} \right) (mRZ\alpha) + \dots \right\}  \,,
\eea
These expressions apply for general $\hat f_i$ and $\hat g_i$ out to subdominant order $mRZ\alpha$ and $(Z\alpha)^2$, and so suffice for modern comparisons with precision measurements. As such they provide a model-independent description of source effects, allowing source effects to be efficiently parameterized when comparing modern measurements \cite{Parthey} with other precisions corrections, such as those of QED.

Finally, we have verified explicitly that these expressions reproduce those in the literature when specialized to the special case where the source is modelled as an explicit charge distribution, and for comparison purposes give expressions for the leading values of $\hat f_i$ and $\hat g_i$ for several simple models considered elsewhere. 

\section*{Acknowledgements}

We thank Paddy Fox, Richard Hill, Bob Holdom, Marko Horbatsch, Friederike Metz, Bernie Nickel, Sasha Penin, Ryan Plestid, Maxim Pospelov, Ira Rothstein, Andrew Tolley and Michael Trott for helpful discussions and Ross Diener, Leo van Nierop, Claudia de Rham and Matt Williams for their help in understanding singular fields and classical renormalization. This research was supported in part by funds from the Natural Sciences and Engineering Research Council (NSERC) of Canada and by a postdoctoral fellowship from the National Science Foundation of Belgium (FWO). Research at the Perimeter Institute is supported in part by the Government of Canada through Industry Canada, and by the Province of Ontario through the Ministry of Research and Information (MRI).  

\appendix

\section{Gamma-matrix conventions}
\label{App:DiracConventions}

When necessary we use the following representation for the tangent-frame gamma matrices:
\be
 \gamma^0 = -i \beta = -i \left(
\begin{array}{cc}
 0 & I  \\
 I &  0      
\end{array}
\right) \,, \quad \gamma_k  = -i \left(
\begin{array}{cc}
 0 & \sigma_k  \\
 -\sigma_k & 0      
\end{array}
\right) \,,
\ee
where $\sigma_k$ are the Pauli matrices,
\be
 \sigma_1 = \left(
\begin{array}{cc}
 0 & 1  \\
 1 &  0      
\end{array}
\right) \,, \quad
 \sigma_2 = \left(
\begin{array}{cc}
 0 & -i  \\
 i &  0      
\end{array}
\right) \quad \hbox{and} \quad
 \sigma_3 =  \left(
\begin{array}{cc}
 1 & 0  \\
 0 & -1      
\end{array}
\right) \,, 
\ee
and $I$ is the 2-by-2 unit matrix. The gamma matrices are defined to satisfy the Dirac algebra $\{ \gamma^\mu, \gamma^\nu \} = 2\, \eta^{\mu\nu}$ where $\eta^{\mu\nu}$ is the inverse Minkowski metric, given (in rectangular coordinates) by $\hbox{diag}(-+++)$. 

Similarly
\be
 \gamma_5 = -i \gamma^0 \gamma^1 \gamma^2 \gamma^3 = \left(
\begin{array}{cc}
 I & 0  \\
 0 & -I      
\end{array}
\right) \,,
\ee
and $\psibar := \psi^\dagger \beta = i \psi^\dagger \gamma^0$. The chirality projection matrices are 
\be
 \gamma_\ssL = \frac12(1 + \gamma_5) \quad \hbox{and} \quad \gamma_\ssR = \frac12(1 - \gamma_5) \quad
 \hbox{so} \quad \psi = \left( \begin{array}{c} \psi_\ssL  \\ \psi_\ssR  \end{array} \right) \,. 
\ee
As usual, the Pauli matrices satisfy 
\be
 \{ \sigma_i, \sigma_j \} = 2\, \delta_{ij} \qquad \hbox{and} \qquad [\sigma_i , \sigma_j] =  2i\,\epsilon_{ijk} \sigma_k \,,
\ee
and so defining $\gamma^{\mu\nu} := \frac12 [ \gamma^\mu, \gamma^\nu]$ we have
\be
 \gamma_{0k} = \frac12 [ \gamma_0 , \gamma_k ] =  \frac12 \left( \begin{array}{cc}
 -2 \sigma_k  & 0  \\
 0 &2 \sigma_k     
\end{array}
\right) = \left( \begin{array}{cc}
 - \sigma_k  & 0  \\
 0 & \sigma_k     
\end{array}
\right)  \,.
\ee 
while 
\be
 \gamma_{jk} = \frac12 [ \gamma_j , \gamma_k ] =  \frac12 \left( \begin{array}{cc}
 [\sigma_j , \sigma_k ] & 0  \\
 0 &[\sigma_j , \sigma_k ]    
\end{array}
\right) = i \epsilon_{jkl} \left( \begin{array}{cc}
 \sigma_l & 0  \\
 0 & \sigma_l   
\end{array}
\right)  \,.
\ee 
Consequently the spin parts of the boost and rotation generators are block-diagonal in this basis, since
\be
 B_j :=- \frac{i}2 \, \gamma_{0j} = \frac{i}2 \left( \begin{array}{cc}
 \sigma_j & 0  \\
 0 & -\sigma_j   
\end{array}
\right)  \quad \hbox{and}\quad
 \Sigma_j :=- \frac{i}4 \, \epsilon_{jkl} \gamma^{kl} = \frac12 \left( \begin{array}{cc}
 \sigma_j & 0  \\
 0 & \sigma_j   
\end{array}
\right)  \,.
\ee

\subsection{Polar coordinates}

Our conventions for spherical polar coordinates $\{r, \theta, \phi \}$ are standard, with (as usual)
\be
 x = r \sin \theta \cos \phi \,, \quad y = r \sin \theta \sin \phi \quad \hbox{and}\quad z = r \cos \theta \,. 
\ee
The differentials therefore satisfy
\be \label{differentials1}
\left( \begin{array}{c} \exd x \\ \exd y \\ \exd z \end{array} \right) =\left(
\begin{array}{ccc} \sin \theta \cos \phi  & \cos \theta \cos \phi &- \sin \phi  \\ \sin \theta \sin \phi  & \cos \theta \sin \phi  & \cos \phi \\ \cos \theta & -\sin \theta &  0   \end{array} \right) 
\left( \begin{array}{c} \exd r \\ r \,\exd \theta \\ r \sin \theta\, \exd \phi \end{array} \right) 
\ee
in terms of which the flat 3D metric is 
\be
 g_{ij} \,\exd x^i \exd x^j = \exd x^2 + \exd y^2 + \exd z^2 = \exd r^2 + r^2 ( \exd \theta^2 + \sin^2 \theta \exd \phi^2) =: (\bfe^r)^2 + (\bfe^\theta)^2 + (\bfe^\phi)^2 \,.
\ee
This last equality defines the normalized frame of basis 1-forms, $\bfe_r$, $\bfe_\theta$ and $\bfe_\phi$, so that an orthonormal frame is given by
\be
 \bfe^r = \exd r \,, \quad \bfe^\theta = r \, \exd \theta \quad \hbox{and} \quad \bfe^\phi = r \sin\theta \, \exd \phi \,.
\ee
We implicitly work in a gauge with $\partial_\mu A^\mu = 0$.  For later use notice the inverse of \pref{differentials1} is
\be
 \left( \begin{array}{c} \exd r \\ r \,\exd \theta \\ r \sin \theta\, \exd \phi \end{array} \right) =\left(
\begin{array}{ccc} \sin \theta \cos \phi  &  \sin \theta \sin \phi  &\cos \theta \\ \cos \theta \cos \phi & \cos \theta \sin \phi  & -\sin \theta \\ - \sin \phi  &  \cos \phi  &  0   \end{array} \right) 
\left( \begin{array}{c} \exd x \\ \exd y \\ \exd z \end{array} \right) \,.
\ee

The radial gamma matrices then are defined by
\bea
 \gamma^\theta &=& \gamma^1 {e_x}^\theta + \gamma^2 {e_y}^\theta + \gamma^3 {e_z}^\theta \nn\\
 &=&   \frac{1}{r} \left[  (\gamma^1 \cos \phi + \gamma^2  \sin \phi)\cos \theta - \gamma^3  \sin \theta \right]\\
 &=& -\frac{i}{ r}\left(
\begin{array}{cc}
 0 & \sigma^\theta  \\
 -\sigma^\theta & 0      
\end{array}
\right) \,, \nn
\eea
with 
\be
 \sigma^\theta := (\sigma_x \cos \phi + \sigma_y \sin \phi) \cos \theta - \sigma_z \sin \theta = \left(
\begin{array}{cc}
 -\sin \theta & e^{-i\phi} \cos \theta  \\
e^{i\phi} \cos \theta & \sin \theta      
\end{array}
\right) \,.
\ee
Similarly 
\bea
 \gamma^\phi &=& \gamma^1 {e_x}^\phi + \gamma^2 {e_y}^\phi + \gamma^3 {e_z}^\phi \nn\\
 &=& \frac{1}{r \sin\theta} \left[- \gamma^1  \sin \phi + \gamma^2  \cos \phi \right]\nn\\
 &=& -\frac{i}{r\sin\theta} \left(
\begin{array}{cc}
 0 & \sigma^\phi  \\
 -\sigma^\phi & 0      
\end{array}
\right) \,, \nn
\eea
with 
\be
 \sigma^\phi := - \sigma_x \sin \phi + \sigma_y \cos \phi  = \left(
\begin{array}{cc}
 0 & -ie^{-i\phi}  \\
ie^{i\phi}  &      0 
\end{array}
\right) \,,
\ee
and (for completeness)
\bea
 \gamma^r &=& \gamma^1 {e_x}^r + \gamma^2 {e_y}^r + \gamma^3 {e_z}^r \nn\\
 &=&  (\gamma^1 \cos \phi + \gamma^2  \sin \phi)\sin \theta + \gamma^3  \cos \theta \\
 &=& -i \left(
\begin{array}{cc}
 0 & \sigma^r  \\
 -\sigma^r & 0      
\end{array}
\right) \,, \nn
\eea
with 
\bea
 \sigma^r &:=& (\sigma_x \cos \phi + \sigma_y \sin \phi) \sin \theta + \sigma_z \cos \theta = \left(
\begin{array}{cc}
 \cos \theta & e^{-i\phi} \sin \theta  \\
e^{i\phi} \sin \theta & -\cos \theta      
\end{array}
\right) \nn\\
&=&  \frac{1}{r} \left( \begin{array}{cc}
 z & x-iy  \\
x+iy  & -z      
\end{array} \right) = \frac{\bfr}{r} \cdot \vec \sigma \,.
\eea

Finally
\be
  \sigma^\theta \sigma^\phi  = -   \sigma^\phi \sigma^\theta=i \left(
\begin{array}{cc}
 \cos \theta &  e^{-i\phi} \sin \theta  \\
e^{i\phi} \sin \theta & - \cos \theta      
\end{array}
\right)  =  i \sigma^r\,,
\ee
and so 
\be
 \gamma^{\theta\phi} = \gamma^\theta \gamma^\phi = - \gamma^\phi \gamma^\theta=  \frac{1}{r^2\sin\theta} \left(
\begin{array}{cc}
 \sigma^\theta\sigma^\phi & 0 \\
 0 & \sigma^\theta\sigma^\phi      
\end{array}
\right) =  \frac{i}{r^2\sin\theta} \left(
\begin{array}{cc}
 \sigma^r & 0 \\
 0 & \sigma^r     
\end{array}
\right)  \,,
\ee
which also implies
\be
 \gamma_{\theta\phi} = g_{\theta\theta} g_{\phi\phi} \gamma^\theta \gamma^\phi =  ir^2\sin\theta \left(
\begin{array}{cc}
 \sigma^r & 0 \\
 0 & \sigma^r     
\end{array}
\right)  \,.
\ee
For future reference notice also that with the convention $\epsilon^{0r\theta\phi} = +\det \, {e_a}^\mu = +1/(r^2 \sin\theta)$ the above imply
\be
 \gamma^{0r} = -\gamma^{r0} = \gamma^0 \gamma^r = \left( \begin{array}{cc} \sigma^r & 0 \\ 0 & -\sigma^r  \end{array} \right)  \quad \hbox{and}
 \quad \gamma_{\theta\phi} \gamma_5 =   ir^2\sin\theta \left( \begin{array}{cc} \sigma^r & 0 \\ 0 & -\sigma^r  \end{array} \right) \,,
\ee
and so
\be
 \gamma^{\mu\nu} = - \frac{i}2 \, \epsilon^{\mu\nu\lambda\rho} \, \gamma_{\lambda\rho} \gamma_5 \,.
\ee

Solutions to the Dirac equation, $(\Dsl + m) \psi = 0$ also solve 
\be \label{KGdirac}
 0 = (\Dsl - m)(\Dsl + m) \psi = (\Dsl^2 - m^2) \psi = \left[ D_\mu D^\mu - m^2 + \frac{i e}{2} \gamma^{\mu\nu} F_{\mu\nu} \right] \psi \,,
\ee
which is the Klein-Gordon equation supplemented by a spin term, whose explicit form is
\be
  + \frac{ie}2 \, \gamma^{\mu\nu} F_{\mu\nu} = + ie \, \gamma^{r0} F_{r0}=  -\frac{iZ\alpha}{ r^2} \,\left(
\begin{array}{cc}
 \sigma^r & 0 \\
 0 & -\sigma^r     
\end{array}
\right)  \,,
\ee
and we use the definition of the fine-structure constant: $\alpha := e^2/(4\pi)$. Once a solution, $\chi$, to \pref{KGdirac} is found, then the corresponding electron-type solution to the Dirac equation $(\Dsl + m )\psi = 0$ is $\psi = (\Dsl - m) \chi$ [and similarly the corresponding positron-type solution to $(\Dsl - m)\psi = 0$ would be $ \psi = (\Dsl + m) \chi$]. 

\subsection{Spinor harmonics}

When solving the Dirac equation we define quantities having definite quantum numbers ($j$ and $j_z$) for $\bfJ$ and $J_z$, leading to the following 2-component spinors
\bea
 U^+_{j\, j_z}(\theta,\phi) &:=& \left[ \begin{array}{c} \sqrt{(j+j_z)/(2j)} \; Y_{j-\frac12 \, j_z - \frac12}(\theta,\phi) \\   \sqrt{(j-j_z)/(2j)} \; Y_{j - \frac12 \, j_z + \frac12}(\theta,\phi)   \end{array}  \right] \nn\\
 \hbox{and} \quad U^-_{j \,j_z}(\theta,\phi) &:=& \left[ \begin{array}{c} \sqrt{(j+1-j_z)/[2(j+1)]} \; Y_{j+\frac12 \, j_z - \frac12}(\theta,\phi) \\  - \sqrt{(j+1+j_z)/[2(j+1)]} \; Y_{j + \frac12 \, j_z + \frac12}(\theta,\phi)   \end{array}  \right] \,.
\eea
Notice that the property $Y_{\ell \,\ell_z}(\pi - \theta, \phi + \pi) = (-)^\ell Y_{\ell \,\ell_z}(\theta, \phi)$ implies parity acts on these combinations oppositely: $\hat \Pi \, U^\pm_{j\, j_z} = (-)^{j\mp \frac12} U^\pm_{j\,j_z}$. Furthermore, notice also that $\sigma^r \, U^\pm_{j\,j_z} = U^\mp_{j\,j_z}$. Indeed the result $\sigma^r \, U^\pm_{j\,j_z} = \eta \, U^\mp_{j\,j_z}$ with $\eta^2 = 1$ is a consequence of the properties $(i)$ $\sigma^r = \hat \bfr \cdot \sigma$ is parity odd; $(ii)$ $(\sigma^r)^2 =1$; and ($iii$) $[ \bfJ, \sigma^r] = 0$, so a direct calculation only determines $\eta = 1$ rather than $\eta = -1$. 

We directly evaluate for the case of most interest: $j = \frac12$. For this purpose we use the explicit forms 
\bea
  &&Y_{00} = \frac{1}{\sqrt{4\pi}} \,, \quad Y_{10} = \sqrt{\frac{3}{4\pi}} \; \cos \theta = \sqrt{\frac{3}{4\pi}} \; \frac{z}{r} \nn\\
  &&Y_{1\pm1} =\mp \sqrt{\frac{3}{8\pi}} \; e^{\pm i \phi} \sin \theta =\mp \sqrt{\frac{3}{8\pi}} \; \frac{x \pm i y}{r}  \,,
\eea
in the definitions of the $U^\pm_{\frac12\, j_z}$ to find
\bea
 U^+_{\frac12\, \frac12}(\theta,\phi) &:=& \left[ \begin{array}{c}  Y_{00}(\theta,\phi) \\  0  \end{array}  \right] =  \frac{1}{\sqrt{4\pi}}  \left[ \begin{array}{c} 1 \\  0  \end{array}  \right] \nn\\
  U^+_{\frac12\, -\frac12}(\theta,\phi) &:=& \left[ \begin{array}{c} 0 \\    Y_{00}(\theta,\phi)   \end{array}  \right] =  \frac{1}{\sqrt{4\pi}}  \left[ \begin{array}{c} 0 \\  1  \end{array}  \right]  \,,
\eea
and 
\bea
 U^-_{\frac12 \,\frac12}(\theta,\phi) &:=& \frac{1}{\sqrt3} \left[ \begin{array}{c} Y_{1 0}(\theta,\phi) \\   -\sqrt2 \; Y_{1 1}(\theta,\phi)   \end{array}  \right] = \frac{1}{\sqrt{4\pi}\; r}  \left[ \begin{array}{c} z \\   x+iy   \end{array}  \right] = \frac{1}{\sqrt{4\pi}}  \left[ \begin{array}{c} \cos \theta \\    e^{i\phi} \sin \theta   \end{array}  \right] \nn\\
 U^-_{\frac12 \,-\frac12}(\theta,\phi) &:=& \frac{1}{\sqrt3} \left[ \begin{array}{c} \sqrt2 \; Y_{1 \, -1}(\theta,\phi) \\  - Y_{1 \, 0}(\theta,\phi)   \end{array}  \right] = \frac{1}{\sqrt{4\pi}\; r}  \left[ \begin{array}{c} x-iy   \\  -z  \end{array}  \right] = \frac{1}{\sqrt{4\pi}}  \left[ \begin{array}{c}  e^{-i\phi} \sin \theta  \\  -\cos \theta  \end{array}  \right] \,,\nn\\
\eea
which are also what is found explicitly by acting on $U^+_{\frac12\,j_z}$ with the explicit matrix
\be
  \sigma^r = \frac{\bfr}{r}\cdot \vec \sigma = \frac{1}{r} \left( \begin{array}{ccc}
  z & x-iy   \\
  x+iy & -z   \end{array} \right) \,.
\ee
Similarly, acting with $\sigma^r$ on $U^-_{\frac12\,j_z}$ gives
\bea
 \sigma^r U^-_{\frac12 \,\frac12}(\theta,\phi) &:=&  \frac{1}{\sqrt{4\pi}\; r^2} \left( \begin{array}{ccc}
  z & x-iy   \\
  x+iy & -z   \end{array} \right)  \left[ \begin{array}{c} z \\   x+iy   \end{array}  \right] =  \frac{1}{\sqrt{4\pi}}  \left[ \begin{array}{c} 1 \\   0  \end{array}  \right] = U^+_{\frac12\,\frac12} \nn\\
 \sigma^r  U^-_{\frac12 \,-\frac12}(\theta,\phi) &:=&  \frac{1}{\sqrt{4\pi}\; r^2}  \left( \begin{array}{ccc}
  z & x-iy   \\
  x+iy & -z   \end{array} \right)  \left[ \begin{array}{c} x-iy   \\  -z  \end{array}  \right]  =  \frac{1}{\sqrt{4\pi}}  \left[ \begin{array}{c} 0  \\  1  \end{array}  \right] = U^+_{\frac12\,-\frac12}   \,,\nn\\
\eea

For later purposes we also evaluate the spatial derivatives explicitly using $\vec\sigma \cdot \nabla =\sigma_k \,\partial_k =  \sigma_x \partial_x + \sigma_y \partial_y + \sigma_z \partial_z $ as well as $\vec\sigma \cdot \nabla f(r) = f'(r) \,\vec\sigma \cdot \nabla r = f'(r) \, \vec\sigma \cdot \bfr /r  = f'(r) \, \sigma^r$. This trivially gives 
\be
  \sigma_k \partial_k \, U^+_{\frac12\,j_z} = 0 \,,
\ee
while
\be
  \sigma_k \partial_k \; U^-_{\frac12\,\frac12} = \frac{2}{r} \; U^+_{\frac12\,\frac12} \quad \hbox{and} \quad
   \sigma_k \partial_k \; U^-_{\frac12\,-\frac12}   =  \frac{2}{r} \; U^+_{\frac12\,-\frac12} \,,
\ee
in agreement with an algebraic evaluation.

\section{Dirac solutions}
\label{App:DiracSolutions}

This appendix collects several exact and approximate solutions to the Dirac equation that are used in the main text.

\subsection{Exterior (Coulomb) solutions}

Bound states for the Dirac equation are found as usual by demanding that the boundary condition (normalizability) at infinity be compatible with the boundary condition at the origin. 

\subsubsection*{Energy eigenvalues}

If the boundary condition at the origin is the usual one (for which we discard the singular solution to the radial equation --- see below) the energy eigenvalues  are
\bea \label{DiracE0}
 \omega_{\ssN} &=& m \left[ 1 + \frac{(Z \alpha)^2}{ \left(N + \zeta \right)^2} \right]^{-1/2} = m \left[ 1 + \frac{(Z \alpha)^2}{ \left[N + \sqrt{\left( j + \frac12 \right)^2 - (Z\alpha)^2} \right]^2} \right]^{-1/2}  \nn\\
 &=&   m \left[ 1 + \frac{(Z \alpha)^2}{ \left(n + \zeta - j - \frac12 \right)^2} \right]^{-1/2}  \,,
\eea
where $j = \frac12, \frac32, \cdots$ and the principal quantum number is defined by $n = N + \left( j + \frac12 \right) = 1,2,3, \cdots$. We define $\zeta = \frac12\sqrt{1 + 4j(j+1) - 4(Z\alpha)^2}$ or
\be \label{zetadef}
  \zeta := \sqrt{\left( j+\frac12 \right)^2 - (Z \alpha)^2} \,,
\ee
so $\zeta \to 1$ as $Z\alpha \to 0$ when $j = \frac12$. This implies
\be \label{useful1}
 \left( j + \frac12 + \zeta \right) \left( j + \frac12 - \zeta \right) = \left( j + \frac12\right)^2 - \zeta ^2  =  \left(Z \alpha \right)^2  \,.
\ee

The standard derivation shows that for $N \ne 0$ (that is, except for $n =  j + \frac12$) each state with fixed $n$ and $j$ comes two-fold degeneracy corresponding to parity $s = \pm$. The most famous example is $N=1$ and $j = \frac12$, which corresponds to $n = 2$ and $j =\frac12$ in which case the degeneracy is between the $2S_{1/2}$ and $2P_{1/2}$ states that get split by the Lamb shift. This two-fold degeneracy does {\em not} occur for $N=0$, corresponding to the $n = j + \frac12$ states like $1S_{1/2}$ (the ground state), $2P_{3/2}$, $3D_{5/2}$ and so on. (Notice that here $S$, $P$ and $D$ do not strictly correspond to specifying $\ell$ but instead give the parity value $s$ for the corresponding state.)

\subsubsection*{Parity Eigenstates}

Normally atomic states are given as parity eigenstates, which involves combining $\psi_\ssL$ and $\psi_\ssR$ since the action of parity is
\be
\cP  \left[ \begin{array}{c}  \psi_\ssL(\theta,\phi) \\ \psi_\ssR(\theta,\phi) \end{array} \right] \cP^{-1} =  \left( \begin{array}{cc}  0 & i \\ i & 0 \end{array} \right)  \left[ \begin{array}{c}  \psi_\ssL(\pi-\theta, \phi + \pi) \\ \psi_\ssR(\pi - \theta, \phi + \pi) \end{array} \right] \,.
\ee

We expect a unique solution for each choice of parity, $j$ and $j_z$ quantum numbers, while the above just relates the radial functions for left- and right-handed fields to one another. The Dirac equation reads
\be 
 -i( D_0 + \sigma_k D_k ) \psi_\ssR + m\, \psi_\ssL = 0 \quad \hbox{and} \quad -i (D_0 - \sigma_k D_k) \psi_\ssL + m\, \psi_\ssR = 0 \,,
\ee
and so
\be \label{diracnabla}
 i \sigma_k \nabla_k \psi_\ssL = \left( \omega + \frac{Z\alpha}{r} \right) \psi_\ssL - m \,\psi_\ssR  \quad \hbox{and} \quad
 i \sigma_k \nabla_k \psi_\ssR = - \left( \omega + \frac{Z\alpha}{r} \right) \psi_\ssR + m \,\psi_\ssL \,.
\ee

To identify the parity eigenstates we expand in terms of the spinor harmonics $U^+$ and $U^-$ of Appendix \ref{App:DiracConventions} and define the radial functions $f(r)$ and $g(r)$ using the following ans\"atze:
\bea
 \psi_\ssL^+ &=& f_+ (r) \,U^+_{j\,j_z}(\theta, \phi) + i g_+(r) \,U^{-}_{j\,j_z}(\theta,\phi) \nn\\
  \hbox{and} \quad
 \psi_\ssR^+ &=& f_+(r) \,U^{+}_{j\,j_z}(\theta,\phi) - i g_+(r) \,U^{-}_{j\,j_z}(\theta,\phi) \nn\\
 \psi_\ssL^- &=& f_- (r) \,U^-_{j\,j_z}(\theta, \phi) + i g_-(r) \,U^{+}_{j\,j_z}(\theta,\phi) \nn\\
  \hbox{and} \quad
 \psi_\ssR^- &=& f_-(r) \,U^{-}_{j\,j_z}(\theta,\phi) - i g_-(r) \,U^{+}_{j\,j_z}(\theta,\phi) \,,
\eea
where the superscript on $\psi$ and subscripts on $f$ and $g$ are the parity eigenlabel $p = \pm$.  Using this in either of \pref{diracnabla} gives the same conditions relating $g$ and $f$. For the parity even states the relations are
\be \label{fgpluseqs}
  f_+' = \left( m + \omega + \frac{Z\alpha}{r} \right) g_+ \quad \hbox{and} \quad
  g_+' + \frac{2g_+}{r} = \left( m - \omega - \frac{Z \alpha}{r} \right) f_+ \,,
\ee
while for parity odd states these relations instead become
\be \label{fgminuseqs}
  g_-' = \left( m - \omega - \frac{Z\alpha}{r} \right) f_- \quad \hbox{and} \quad
  f_-' + \frac{2f_-}{r} = \left( m + \omega + \frac{Z \alpha}{r} \right) g_- \,,
\ee
as used in the main text.

\subsubsection*{Coulomb-Dirac solutions}

To solve the radial Dirac equations, \eqref{fgpluseqs} and \eqref{fgminuseqs}, for general radius we introduce the two functions
\begin{align}
 \begin{aligned} \label{Q_1Q_2Ansatz}
  Q_1 &= \frac12 e^{\rho/2} \rho^{1-\zeta} \left(\frac{f}{\sqrt{m+\omega}} - \frac{g}{\sqrt{m-\omega}}\right)\\
  Q_2 &= \frac12 e^{\rho/2} \rho^{1-\zeta} \left(\frac{f}{\sqrt{m+\omega}} + \frac{g}{\sqrt{m-\omega}}\right)\\
 \end{aligned}
\end{align}
where $\rho = 2 \kappa r$ and $\kappa = \sqrt{m^2-\omega^2}$. Some manipulation shows that these satisfy the following first-order linear ODEs
\begin{align}
 \begin{aligned} \label{Q_1Q_22ndordereq}
  \rho Q_1'' + (2\zeta+1-\rho) Q_1' - (\zeta -\frac{Z\alpha \omega}{\kappa})Q_1 &= 0\\
  \rho Q_2'' + (2\zeta+1-\rho) Q_2' - (\zeta +1 -\frac{Z\alpha \omega}{\kappa})Q_2 &= 0 \,,\\
 \end{aligned}
\end{align}
which hold for {\em either} sign of the parity quantum number. The parameter $\zeta$ is as defined in \pref{zetadef}. The most general solutions to these equations are given as linear combinations of confluent hypergeometric functions $\cM(a,b;\rho) =  1 + (a/b)\rho + \cdots$, thereby introducing a total of four integration constants. 

The Dirac equation imposes two relations between the four constants. Hence, we can express the solutions $Q_1$ and $Q_2$ as
\bea
 \label{Q_1Q_2solAC}
  Q_1 &=& A\, \cM\left[ \zeta -\frac{Z\alpha \omega}{\kappa}, 2\zeta+1;\rho\right] + C\, \rho^{-2\zeta} \cM\left[-\zeta -\frac{Z\alpha \omega}{\kappa},-2\zeta+1;\rho\right] \\
  Q_2 &=& -A\, \left( \frac{\zeta-{Z\alpha \omega}/{\kappa}}{K-{Z\alpha m}/{\kappa}} \right) \, \cM \left[\zeta -\frac{Z\alpha \omega}{\kappa}+1,2\zeta+1;\rho\right] \nn\\
  && \qquad\qquad +C\,\left( \frac{\zeta+{Z\alpha \omega}/{\kappa}}{K-{Z\alpha m}/{\kappa}} \right)\, \rho^{-2\zeta} \cM\left[ -\zeta -\frac{Z\alpha \omega}{\kappa}+1,-2\zeta+1;\rho\right] \,,
\eea
where $K = \mp (j+\frac12)$ for states with parity $\pm 1$. $A$ and $C$ are the two remaining integration constants, and are chosen so that the function multiplying $A$ is bounded as $\rho \to 0$ while the function multiplying $C$ diverges there. 

The corresponding expressions for $f$ and $g$ are then given by
\bea \label{fACgen}
    f &=& \sqrt{m+\omega}\,  e^{-\rho/2} \rho^{\zeta-1} \left\{ A \, \cM \left[ \zeta -\frac{Z\alpha \omega}{\kappa}, 2\zeta+1; \rho \right] +  C \rho^{-2\zeta} \cM \left[ -\zeta -\frac{Z\alpha \omega}{\kappa}, -2\zeta+1; \rho \right] \right. \nn\\
    &&\qquad\qquad\qquad\qquad  -  A \left(\frac{\zeta-{Z\alpha \omega}/{\kappa}}{K-{Z\alpha m}/{\kappa}}\right) \cM \left[ \zeta -\frac{Z\alpha \omega}{\kappa}+1, 2\zeta+1; \rho \right] \\
    &&\qquad\qquad \qquad\qquad\qquad \left. +C \left(\frac{\zeta+{Z\alpha \omega}/{\kappa}}{K-{Z\alpha m}/{\kappa}}\right) \rho^{-2\zeta} \cM \left[ -\zeta -\frac{Z\alpha \omega}{\kappa}+1, -2\zeta+1; \rho \right] \right\} \,, \nn
\eea
and
\bea \label{gACgen}
    g &=& -\sqrt{m-\omega}\,  e^{-\rho/2} \rho^{\zeta-1} \left\{ A \, \cM \left[ \zeta -\frac{Z\alpha \omega}{\kappa},2\zeta+1;\rho \right] + C \rho^{-2\zeta} \cM \left[ -\zeta -\frac{Z\alpha \omega}{\kappa},-2\zeta+1;\rho\right] \right. \nn\\
    &&\qquad\qquad\qquad\qquad  + \left.  A \left( \frac{\zeta-{Z\alpha \omega}/{\kappa}}{K-{Z\alpha m}/{\kappa}} \right) \cM \left[\zeta -\frac{Z\alpha \omega}{\kappa}+1,2\zeta+1;\rho\right] \right. \\
    &&\qquad\qquad \qquad\qquad\qquad \left.- C \left( \frac{\zeta+{Z\alpha \omega}/{\kappa}}{K-{Z\alpha m}/{\kappa}}\right) \rho^{-2\zeta} \cM \left[-\zeta -\frac{Z\alpha \omega}{\kappa}+1,-2\zeta+1;\rho\right] \right\} \,.\nn
\eea

Normalisation of the state for $\rho \rightarrow \infty$ demands $A$ and $C$ must be related by
\begin{equation} \label{ACGammas}
 \frac{A}{C} = -\, \frac{\Gamma(1-2\zeta)}{\Gamma(1+2\zeta)}\,\frac{\Gamma(\zeta-{Z\alpha \omega}/{\kappa})}{\Gamma(-\zeta-{Z\alpha \omega}/{\kappa})} 
\end{equation}
which follows from the the large-$\rho$ form of the confluent hypergeometric functions $\cM$. When $C = 0$ this condition reproduces the energy eigenvalue given in \pref{DiracE0}. Alternative boundary conditions at $r \to 0$ change the bound state energy levels (and any other physical implications) entirely by changing what they imply for $A/C$. 

As the above formulae attest, such alternative boundary conditions governing $A/C$ can be imposed by demanding that the ratio $f/g$ take a specific value at a particular radius $r=\epsilon$. (For instance, for particles orbiting a known charge distribution that extends out to radius $r = R$, it is continuity of the internal with the external solution at $r = R$ that imposes the required condition:
\begin{equation} \label{fgmatch}
 \frac{g_{\text{out}}(R,K)}{f_{\text{out}}(R,K)} = \frac{g_{\text{in}}(R,K)}{f_{\text{in}}(R,K)}
\end{equation}
where $f_{\text{out}}$ and $g_{\text{out}}$ are the Coulomb solutions described above, valid for $r > R$, and $f_{\text{in}}$ and $g_{\text{in}}$ are given by the solving the Dirac equation for the charge distribution for $r\leq R$. The next sections provide several representative solutions for simple charge distributions.

\subsection{Interior solutions for given charge distributions}

This section collects several simple solutions appropriate to the interior for several kinds of charge distributions, and gives the approximate series solutions in the general case.

\subsubsection{Charged-shell model}

In this case consider an exactly solvable model of a charge distribution against which later results can be compared. The model assumes a charge distribution that makes up a spherical shell, with surface density $\sigma$. That is,
\be
 \rho = \sigma \, \delta(r - R) = \frac{Ze}{4\pi R^2} \, \delta (r-R)\,
\ee
where $R$ is the radius of the shell, and the second equality assumes the total charge is $Z e$. The corresponding electromagnetic potential found by integrating Maxwell's equations then is
\be
  A^0 = \frac{Ze}{4\pi r} \quad \hbox{if $r > R$} \qquad \hbox{and} \qquad A^0 = \frac{Ze}{4\pi R} \quad \hbox{if $r < R$} \,.
 \ee
 
The Dirac equation outside the shell is therefore sees only the Coulomb potential and so is the one whose solutions are given above. The solution inside the shell is essentially the free Dirac equation, though in the presence of a nonzero constant $A^0$. That is, it is equivalent to \pref{KGdirac}, which now reads
\be  
 0 = (\Dsl - m)(\Dsl + m) \psi  = \left[ D_\mu D^\mu - m^2 + \frac{i e}{2} \gamma^{\mu\nu} F_{\mu\nu} \right] \psi =  \left[ D_\mu D^\mu - m^2  \right] \psi \,,
\ee
where the spatial derivatives are $D_i = \partial_i$ while the time derivative (acting on an energy eigenstate) is 
\be
 D_0 = \partial_t +ie A_0 = -i \left( \omega + \frac{Z\alpha}{4\pi R} \right) \,.
\ee

This has as solutions the usual spherical Bessel functions
\be
   A\, j_\ell(kr) + B\, y_\ell (kr) \,,
\ee
and $B=0$ if we demand $R$ be bounded at $r = 0$. Specializing to $j = \frac12$ the appropriate solutions are $f_+ = A_+ j_0(kr)$, $f_- = B_- j_1(kr)$, $g_+ = B_+ j_1(kr)$ and $g_- = A_- j_0(kr)$. Since $f_+'$ and $g_-'$ vanish at the origin it follows that $g_+$ and $f_-$ must vanish there and this is automatic because these only involve $\ell = 1$. When evaluated at $r = R$ then 
\be
  \frac{g_+(R)}{f_+(R)} = \left( \frac{B_+}{A_+} \right) \frac{j_1(kR)}{j_0(kR)} \,,
\ee
and 
\be
 \frac{f_-(R)}{g_-(R)} = \left( \frac{B_-}{A_-} \right) \frac{j_1(kR)}{j_0(kR)} \,.
\ee 

Finally, the Dirac equation says $f_+' = (m+W) g_+$ and $g_-' = (m-W) f_-$ where $W = \omega + Z\alpha/R$. Using 
\be
  j_0(x) = \frac{\sin x}{x} \simeq 1 + \cO(x^2) \quad \hbox{and} \quad
  j_1(x) = \frac{\sin x}{x^2} - \frac{\cos x}{x} \simeq \frac{x}{3} + \cO(x^3)\,,
\ee
so $j_0'(x) = - j_1(x)$ we find $f_+' = (m+W)g_+$ implies $-k A_+ = (m+W)B_+$ and $g_-' = (m-W) f_-$ implies $-kA_- = (m-W) B_-$. This allows the boundary condition to be written
\bea
  \frac{g_+(R)}{f_+(R)} &=& - \left( \frac{k}{m+W} \right) \frac{j_1(kR)}{j_0(kR)} = - \sqrt{\frac{W-m}{W+m}} \left[ \frac{\sin(kR) - kR \cos(kR)}{kR\sin(kR)}  \right] \nn\\
   &=& - \frac{1}{3} (W-m) R \left[ 1 + \frac{(kR)^2}{15} + \frac{2(kR)^4}{315} + \cdots \right] \,,
\eea
where we use $(\sin x - x \cos x)/(x \sin x) = \frac13 \, x + \frac{1}{45} \, x^3 + \frac{2}{945} \, x^5 + \cdots$.

To make contact with the series for in powers of $(Z\alpha)^2$ and $mRZ\alpha$ we evaluate at a bound-state energy and use
\bea
  (kR)^2 &=& \Bigl[ (\omega + m)R + Z\alpha \Bigr] \Bigl[ (\omega - m)R + Z\alpha \Bigr] \nn\\
  &\simeq& (2mR + Z\alpha) Z\alpha + \cO[(mRZ\alpha)^2 \; \hbox{or} \; (Z\alpha)^3 mR]\,, \nn
\eea
and 
\be
  (W - m)R = (\omega -m)R + Z\alpha \simeq - \frac{1}{2n^2}(Z\alpha)^2 mR + Z\alpha = Z\alpha \left[1 - \frac{mRZ\alpha}{2n^2} + \cO[(Z\alpha)^3mR ]\right] \,,
\ee
so that
\be
  \frac{g_+(R)}{f_+(R)} \simeq  -\frac{Z\alpha}3 \left[ 1 +\left( \frac{2}{15} - \frac{1}{2n^2} \right) (mRZ\alpha) + \frac{(Z\alpha)^2 }{15} + \cdots \right]   \,,
\ee
which drops terms in the brackets that are of order $mR(Z\alpha)^3$, $(mRZ\alpha)^2$ and $(Z\alpha)^4$.

Similarly, for the parity-odd case
\bea
 \frac{f_-(R)}{g_-(R)} &=&  -\left( \frac{k}{m-W} \right) \frac{j_1(kR)}{j_0(kR)}  =+\sqrt{\frac{W+m}{W-m}} \left[  \frac{\sin(kR) - kR \cos(kR)}{kR\sin(kR)} \right] \nn\\
 &=&  \frac{(W+m)R}{3}  \left[ 1 + \frac{(kR)^2}{15} + \frac{2(kR)^4}{315} + \cdots \right] \,,
\eea 
and so again using the bound-state energy and the above approximate expressions we have 
\bea
  \sqrt{\frac{m-\omega}{m+\omega}} \left[ \frac{f_-(R)}{g_-(R)} \right]  &\simeq& \left( \frac{Z\alpha}{2n} \right) \frac{1}3 \left( 2 mR + Z\alpha \right) \left[ 1 + \frac{Z\alpha}{15} (2mR+Z\alpha) + \cdots \right]   \nn\\
   &\simeq&   \frac1{3n} (mRZ\alpha) + \frac{2}{45} (mRZ\alpha)^2 + \frac{(Z\alpha)^2}{6n} + \cO[mR(Z\alpha)^3;\; (Z\alpha)^4]   \,.
\eea 

These imply $g_+/f_+ \simeq - \frac13 (W - m)R \simeq - \frac13 (Z\alpha)$ in the parity-even case for both the nonrelativistic and relativistic limits, while   $f_-/g_- \simeq  \frac23 \, mR$ in the nonrelativistic limit ($mR \gg Z\alpha$) while in the relativistic limit (for which $Z \alpha/R \gg \omega \simeq m$) we instead find $f_-/g_-  \simeq  \frac13 \, Z \alpha$. 

\subsubsection*{Expansion coefficients}

For comparison with the results for other charge distributions for use in the main text it is useful to quote the above results in terms of parameters $\hat g_i$ and $\hat f_i$ appearing in the expansion
\be
 \left( \frac{g_+}{f_+} \right)_{r=R} = Z\alpha \Bigl[ \hat g_1 + \hat g_2 (mRZ\alpha) + \hat g_3 (Z\alpha)^2 + \cdots \Bigr] \,,
\ee
and
\be
 \sqrt{\frac{m-\omega}{m+\omega}} \left( \frac{f_-}{g_-} \right)_{r=R} = \frac{1}{2n} \Bigl[ \hat f_1 (mRZ\alpha) + \hat f_2 (mRZ\alpha)^2 + \hat f_3 (Z\alpha)^2 + \cdots \Bigr] \,.
\ee
With these definitions the above calculation shows that the charged shell predicts for the parity-even state we have
\be
  \hat g_1 = - \, \frac13 \,, \qquad \hat g_2 = - \frac{2}{45} + \frac{1}{6n^2}  \qquad \hbox{and} \qquad \hat g_3 = - \,\frac{1}{45} \,,
\ee
while for the parity-odd state the parameters are
\be
  \hat f_1 = + \frac23 \,, \qquad \hat f_2 = +  \frac{2}{45} \qquad \hbox{and} \qquad \hat f_3 = + \frac{1}{3}  \,.
\ee

\subsubsection{General charge distribution}

Next evaluate the interior solution for a general distribution $\rho(r)$ for $r \le R$ by evaluating as a series in $kR$. This is generally sufficient since $kR \simeq MRZ\alpha$ or $Z\alpha$ in the cases $mR \gg Z\alpha$ and $mR \ll Z\alpha$. The goal will be to determine $f/g$ at $r=R$ as a function of the first few derivatives of $\rho$ at $r = 0$. 

To this end assume a charge distribution of the form
\be
 \rho = \rho(r) \qquad \hbox{with} \qquad \rho(R) = 0 \quad \hbox{for $r \ge R$}\,,
\ee
where $R$ is the radius of the distribution and 
\be
 4\pi\int_0^\infty \exd r \, r^2 \rho(r) = Z e \,. 
\ee
The corresponding electromagnetic potential satisfies $\bfE = - \nabla A^0$ and so $\nabla \cdot \bfE = - \nabla^2 A^0 = \rho$ and so 
\be
  \nabla^2 A_0 = \frac{1}{r^2} \partial_r \Bigl( r^2 \partial_r A_0 \Bigr) = \rho 
\ee
and so
\be
  A^0 = \frac{Ze}{4\pi r} \quad \hbox{if $r > R$} \,.
\ee

For $r < R$ we use dimensionless variable $u = r/R$ so $A_0(u)$ satisfies
\be
 \frac{1}{u^2} \Bigl( u^2 A_0' \Bigr)' = R^2 \, \rho \,,
\ee 
and if we demand that $\rho$ and $A_0$ must be analytic at $u = 0$ we may demand $\rho(-u) = \rho(u)$ (and similarly for $A_0(u)$) and so write (with a small abuse of notation)
\bea \label{Seriesexp}
  \rho(u) &=& \frac{3Ze}{4\pi R^3} \Bigl[ \rho_0 + \rho_2 \, u^2 +  \rho_4 \, u^4 + \cdots \Bigr] \nn\\
  A_0(u) &=& A_0(0) + A_2 \, u^2 + A_4 \, u^4 + \cdots \,.
\eea
Note that the coefficients $\rho_{2k}$ are not completely independent of each other, since the charge density must satisfy $Ze = \int \exd^3x\, \rho(r)$, and so we must have
\begin{equation}
	\label{eq:rho_constraint}
	\frac 13 = \sum_{k=0}^\infty \frac{\rho_{2k}}{2k + 3}.
\end{equation}
Inserting \pref{Seriesexp} into the Maxwell equation leads to
\be
  6A_2 + 20 A_4 u^2 + \cdots + k(k+1) A_k u^{k-2} + \cdots = \frac{3Ze}{4\pi R} \Bigl[ \rho_0 + \rho_2 u^2 + \cdots + \rho_k u^k + \cdots \Bigr] \,,
\ee
and so 
\be
  A_2 = \frac{Ze\rho_0 }{8\pi R} \;, \qquad  A_4 = \frac{3Ze\rho_2}{80\pi R} \qquad \hbox{and} \qquad  A_k = \frac{3Ze\rho_{k-2}}{4\pi k(k+1) R} \,,
\ee
while continuity at $r=R$ demands 
\be
 A_0(0) + A_2 +  A_4 + \cdots = - \frac{Ze}{4\pi R} \,,
\ee
and so
\bea
 eA_0(r) &=& eA_0(0) + \frac{Z\alpha}{ R} \left[  \frac{\rho_0}{2} \, u^2 + \frac{3\rho_2}{20} \; u^4 + \cdots + \frac{3\rho_{k-2}}{k(k+1)} \; u^{k} + \cdots \right]\nn\\
 &=& \frac{Z\alpha}{ R} \left[ -1 + \frac{\rho_0}{2} \Bigl( u^2 - 1 \Bigr) + \frac{3\rho_2}{20} \Bigl( u^4 - 1 \Bigr) + \cdots + \frac{3\rho_{k-2}}{k(k+1)} \Bigl( u^k - 1 \Bigr) + \cdots \right]\,,
\eea
where $u = r/R$. These identify the parameters --- {\em i.e.} $A_0(0)$, $\rho_0$, $\rho_2$ and so on --- that govern the leading form of the interior solutions to the Dirac equation.  

We now solve the Dirac equation explicitly. The solution outside the shell is sees only the Coulomb potential and so is the one  given in earlier appendices. The solution inside the shell we solve in the presence of the above nonzero potential $A_0(r)$, perturbatively in $u$. 

\subsubsection*{Parity-even states}

For parity-even states the functions $f_+$ and $g_+$ satisfy \pref{fgpluseqs}, which reads
\be \label{fgpluseqs2}
  \partial_r f_+ = \Bigl[ m + \omega -eA_0(r) \Bigr] g_+ \quad \hbox{and} \quad
  \partial_r g_+ + \frac{2g_+}{r} = \Bigl[ m - \omega +eA_0(r) \Bigr] f_+ \,,
\ee
so in terms of $u=r/R$ we find
\bea \label{feq+}
   f_+' &=& R \Bigl[ m + \omega -eA_0(u) \Bigr] g_+ \\
    &=& \left\{  (m+\omega)R -eA_0(0) R- Z\alpha\left[\left( \frac{\rho_0}{2} \right) u^2 +\left(\frac{3\rho_2}{20} \right) u^4 + \cdots \right]\right\} g_+ \,,\nn
\eea
and 
\bea \label{geq+}
    \left( g_+' + \frac{2g_+}{u} \right) &= & R\Bigl[ m - \omega +eA_0(u) \Bigr] f_+ \\
    &=&  \left\{  (m-\omega)R + eA_0(0)R + Z\alpha \left[ \left( \frac{\rho_0}{2}\right)u^2 + \left( \frac{3\rho_2}{20} \right) u^4 + \cdots \right]\right\} f_+\,.\nn
\eea

Writing
\bea
  f_+ &=& \mff^+_0 + \frac12 \, \mff^+_2 \, u^2 + \frac14 \, \mff^+_4 \, u^4 + \cdots \nn\\
  g_+ &=& \mfg^+_1 \, u + \frac13 \, \mfg^+_3 \, u^3 + \frac15 \, \mfg^+_5 \, u^5 + \cdots \,,
\eea
then \pref{feq+} implies
\bea
  \mff^+_2u +  \mff^+_4 u^3 + \cdots  &=& \left\{  (m+\omega)R -eA_0(0)R - Z\alpha \left[ \left( \frac{\rho_0}{2} \right)u^2+ \left( \frac{3\rho_2}{20} \right) u^4 + \cdots  \right]\right\}\nn\\
  && \qquad\qquad\qquad\qquad\qquad\qquad \times \left[ \mfg^+_1 u + \frac13 \mfg^+_3 u^3 + \cdots \right] \,,
 \eea
and so
\bea
  \mff^+_2 &=&  \Bigl[  (m+\omega)R -eA_0(0)R \Bigr] \mfg^+_1 = M_+ \mfg^+_1 \nn\\
 \mff^+_4 &=&   \Bigl[  (m+\omega)R -eA_0(0)R \Bigr] \frac{\mfg^+_3}3 -  \left( \frac{Z\alpha \,\rho_0}{2}   \right) \mfg^+_1 =  \left( \frac{M_+}3 \right) \mfg^+_3 -  \left( \frac{Z\alpha \,\rho_0}{2}   \right) \mfg^+_1 \nn\\
  \mff^+_6 &=&   \Bigl[  (m+\omega)R -eA_0(0)R \Bigr] \frac{\mfg^+_5}5 -  \frac{Z\alpha}{2} \left( \frac{\rho_0\, \mfg^+_3}{3} +  \frac{3\rho_2\, \mfg^+_1}{10} \right) \,,
\eea
and so on, where we define
\be
  M_\pm :=    \Bigl[ m \pm \Bigl( \omega - eA_0(0)\Bigr)\Bigr] R \,.
\ee

Similarly \pref{geq+} implies
\bea
    3\mfg^+_1 + \frac53 \, \mfg^+_3 u^2 + \frac75 \, \mfg^+_5 u^4 + \cdots &=& \left\{ M_- + Z\alpha \left[ \left( \frac{\rho_0}{2} \right)u^2 + \left( \frac{3\rho_2}{20}\right)  u^4 + \cdots\right]\right\} \nn\\
  && \qquad\qquad\qquad\qquad\qquad\qquad \times  \left[ \mff^+_0 + \frac12 \mff^+_2 u^2 + \cdots \right]   
\eea
and so
\bea
  \mfg^+_1 &=& \left( \frac{M_- }3  \right) \mff^+_0 \nn\\
  \mfg^+_3  &=& \frac3{10}  \Bigl( M_- \;\mff^+_2 +  Z\alpha \,\rho_0 \Bigr) \mff^+_0 \\
  \mfg^+_5  &=& \frac5{7} \left[  \left( \frac{M_- }4 \right) \mff^+_4+ \left( \frac{Z\alpha \,\rho_0}{4} \right) \mff^+_2 + \left( \frac{3Z\alpha\, \rho_2}{20}  \right) \, \mff^+_0 \right] \,,\nn
\eea
and so on.

These equations fix all coefficients in terms of the unknown normalization $\mff^+_0$ as well as $A_0(0)$ and the $\rho_i$ which are assumed to be known.  The series for the solution at $r=R$ then takes the form
\be
 f_+ (R) = \mff^+_0 \left[ 1 + \frac{\mff^+_2}{2 \, \mff^+_0}  + \frac{\mff^+_4}{4\, \mff^+_0} + \cdots \right] \qquad \hbox{and} \qquad
 g_+(R) = \mff^+_0 \left[  \frac{\mfg^+_1}{\mff^+_0} +  \frac{\mfg^+_3}{3\, \mff^+_0} +  \frac{\mfg^+_5}{5\, \mff^+_0} + \cdots \right]  \,,
\ee
where
\bea 
  \frac{\mfg^+_1}{\mff^+_0} &=&  \frac{M_-}3   \nn\\
    \frac{\mff^+_2}{2\mff^+_0} &=&   M_+ \left( \frac{\mfg^+_1}{2\mff^+_0} \right) = \frac{M_+M_-}6 = -\frac16 (k_0R)^2  \nn\\
      \frac{\mfg^+_3}{3\mff^+_0} &=& \left( \frac{M_-}5 \right) \frac{\mff^+_2}{2\mff^+_0} + \frac{Z\alpha \,\rho_0}{10} = \frac{Z\alpha \,\rho_0}{10}  + \frac{M_+M_-^2}{30}    = \frac{Z\alpha \,\rho_0}{10}  - \frac{M_-}{30} (k_0R)^2   \nn \\
      \frac{\mff^+_4}{4\mff^+_0} &=&    \left( \frac{M_+}{12} \right) \frac{\mfg^+_3}{\mff^+_0} -  \left( \frac{Z\alpha \,\rho_0}{8}   \right) \frac{\mfg^+_1}{\mff^+_0}   
  =  \frac{Z\alpha \,\rho_0}{8} \left( \frac{M_+}5 -  \frac{M_-}3 \right) + \frac{(k_0R)^4}{120} \\
      \frac{\mfg^+_5}{5\mff^+_0} &=&    \left( \frac{M_-}{7} \right) \frac{\mff^+_4}{4\mff^+_0} +  \left( \frac{Z\alpha \,\rho_0}{14}   \right) \frac{\mff^+_2}{2\mff^+_0} +  \frac{3Z\alpha \,\rho_2}{140}  \nn\\
  &=&    -  \frac{Z\alpha \,\rho_0 \, M_-^2}{168}  +  \frac{M_-}{840} (k_0R)^4 -  \frac{13Z\alpha \,\rho_0}{840} (k_0R)^2   +  \frac{3Z\alpha \,\rho_2}{140}  \nn\\
    \frac{\mff^+_6}{6\mff^+_0} &=&  \left( \frac{M_+}6 \right)   \frac{\mfg^+_5}{5\mff^+_0} -  \left( \frac{Z\alpha \,\rho_0}{12} \right)  \frac{\mfg^+_3}{3\mff^+_0} + \left( \frac{ Z\alpha \, \rho_2}{40} \right) \frac{\mfg^+_1}{\mff^+_0} \nn\\
    &=&    \frac{Z\alpha \,\rho_2}{40} \left(\frac{M_+}{7} + \frac{M_-}3 \right)  -  \frac{(Z\alpha \,\rho_0)^2}{120}+   \frac{Z\alpha \,\rho_0}{5040}(k_0R)^2 \Bigl( 19 M_- - 13 M_+ \Bigr) -  \frac{(k_0R)^6}{5040}     \nn
\eea
and so on. These last equalities define
\be
  k^2_0 := \Bigl[ \omega - eA_0 (0) \Bigr]^2 - m^2  \qquad \hbox{so that} \qquad
  (k_0R)^2 = - M_+M_- \,,
\ee
and because $M_- \sim \cO(Z\alpha)$ and $M_+ \sim \cO[mR + Z\alpha]$ we see that the expansion is controlled by powers of $mRZ\alpha$ and $(Z\alpha)^2$.

The boundary condition of interest in this case is $g_+(R)/f_+(R)$ which is given by
\bea
  \left( \frac{g_+}{f_+} \right)_{r=R} &=& \frac{\mfg^+_1  + \frac13 \mfg^+_3 + \frac15 \mfg^+_5  + \cdots }{\mff^+_0 +\frac12 \mff^+_2+ \frac14 \mff^+_4 + \cdots } \\
  &\simeq& \left[  \frac{\mfg^+_1}{\mff^+_0}   +  \frac{\mfg^+_3}{3\mff^+_0} +  \frac{\mfg^+_5}{5\mff^+_0} + \cdots  \right]\left[1  - \left(   \frac{\mff^+_2}{2\mff^+_0} + \frac{\mff^+_4}{4\mff^+_0} + \frac{\mff^+_6}{6\mff^+_0} +\cdots\right) + \cdots \right] \,.\nn
\eea
Consequently
\bea
  \left( \frac{g_+}{f_+} \right)_{r=R}    &=& \left\{  \frac{M_-}3\left[ 1   - \frac{(k_0R)^2 }{10} \right]  +\frac{Z\alpha \,\rho_0}{10}  \left[ 1 -  \frac{5M_-^2}{84}   -  \frac{13(k_0R)^2}{84}   \right]  +  \frac{3Z\alpha \,\rho_2}{140} \Bigl[ 1 + \cdots \Bigr] + \cdots\right\} \nn\\
  && \times \left\{1 +\frac16 (k_0R)^2+\frac{Z\alpha \,\rho_0}{8} \left( \frac{M_-}3 -  \frac{M_+}5 \right)  - \frac{Z\alpha \,\rho_2}{40} \left(\frac{M_+}{7} + \frac{M_-}3 \right)  +  \frac{(Z\alpha \,\rho_0)^2}{120} + \cdots  \right\}  \nn\\
  &=&  \frac{M_-}3\left[ 1  + \frac{(k_0R)^2 }{15} \right]  +\frac{Z\alpha \,\rho_0}{10}  \left[ 1 +  \frac{5M_-^2}{63}  +  \frac{2(k_0R)^2}{21}  \right]   +\frac{(Z\alpha \,\rho_0)^2}{8} \left( \frac{M_-}{18} -  \frac{M_+}{50} \right) \nn\\
  && \qquad\qquad\qquad\qquad\qquad\qquad +  Z\alpha \,\rho_2 \Bigl[ 1 + \cdots \Bigr] + \cdots  \,. 
\eea
where $(k_0R)^2 = - M_+M_-$ with $M_- \sim \cO(Z\alpha)$ and $M_+ \sim \cO[mR + Z\alpha]$ and drop any terms that are suppressed by more than just $mRZ\alpha$ or $(Z\alpha)^2$ relative to the leading term. 

Notice in particular that higher coefficients $\rho_i$ enter suppressed only by $Z\alpha$. We now show that these terms of order $Z\alpha$ sum to give the result required to have the energy shift be controlled by the mean-square charge distribution 
\be
  r_p^2 := \frac{1}{Ze} \int \exd^3x \, r^2 \rho(x) = 3R^2 \sum_{k=0}^\infty \frac{\rho_{2k}}{2k+5} \,. 
\ee
To see if this is so we track these terms explicitly using
\be
 \frac{\mfg^+_k}{k\mff^+_0} = \frac{3 Z\alpha \,\rho_{k-3}}{k(k-1)(k+2)} + \hbox{(other terms)} \qquad \hbox{for $k = 3,5,7,\cdots$.}
\ee
The leading contribution to $g_+/g_-$ then is
\bea
 \frac{g_+}{f_+} &=& \frac{M_-}3 +  Z\alpha \sum_{k=0}^\infty \frac{3\rho_{2k}}{(2k+2)(2k+3)(2k+5)} + \cdots \nn\\
 &=&  \frac{eA_0(0)R}3 + Z\alpha \sum_{k=0}^\infty \frac{3\rho_{2k}}{(2k+2)(2k+3)(2k+5)} + \cdots
\eea
so using
\be
 eA_0(0)R = -Z\alpha \left[ 1+ \sum_{k=0}^\infty \frac{3 \rho_{2k}}{(2k+2)(2k+3)}  \right]\,,
\ee
we have
\bea
 \frac{g_+}{f_+} &=& Z\alpha \left[-\frac13 + \sum_{k=0}^\infty \left(\frac{3\rho_{2k}}{(2k+2)(2k+3)(2k+5)} -  \frac{ \rho_{2k}}{(2k+2)(2k+3)}  \right)   \right] + \cdots \nn\\
 &=& Z\alpha \left[-\frac13 - \sum_{k=0}^\infty \frac{\rho_{2k}}{(2k+3)(2k+5)}  \right] + \cdots
\eea
This contributes to the effective coupling $h_\text{eff}^+$ the amount
\bea \label{eq:heff_genChargeP}
	h_\text{eff}^+ &\approx& 2\pi Z\alpha R^2 \left\{ 1 + \frac 2 {Z\alpha} \left( \frac {g_+}{f_+} \right)\right\} 
	= 2\pi Z\alpha R^2 \left\{ \frac 1 3 - 2\sum_{k=0}^\infty \frac{\rho_{2k}}{(2k + 3)(2k + 5)} \right\} \nn \\
	   &=& 2\pi Z\alpha R^2 \sum_{k=0}^\infty \frac{\rho_{2k}}{2k + 3}\left\{1 - \frac{2}{2k + 5} \right\}
	   = 2\pi Z\alpha R^2 \sum_{k=0}^\infty \frac{\rho_{2k}}{2k + 5} \\
	   &=& \frac {2\pi} 3 Z\alpha\, r_p^2 \,.\nn  
\eea

\subsubsection*{Parity-odd states}

For parity-odd states the functions $f_-$ and $g_-$ satisfy \pref{fgminuseqs}, which reads
\be \label{fgminuseqs2}
  \partial_r g_- = \Bigl[ m - \omega +eA_0(r) \Bigr] f_- \quad \hbox{and} \quad
  \partial_r f_- + \frac{2f_-}{r} = \Bigl[ m + \omega -eA_0(r) \Bigr] g_- \,,
\ee
which has the same form as did the parity-even case if we make the replacements $f_+ \leftrightarrow g_-$, $f_- \leftrightarrow g_+$ and $\omega - eA_0 \leftrightarrow - (\omega - eA_0)$. This implies the solutions have the same form with $\mfg^\pm_i \leftrightarrow \mff^\mp_i$ as well as $M_+ \leftrightarrow M_-$ and $\rho_i \leftrightarrow - \rho_i$. 

Consequently for parity-odd states we have 
\bea
  \left( \frac{f_-}{g_-} \right)_{r=R}  &=& \frac{\mff^-_1  + \frac13 \mff^-_3 + \frac15 \mff^-_5  + \cdots }{\mfg^-_0 +\frac12 \mfg^-_2+ \frac14 \mfg^-_4 + \cdots }\nn\\
  &\simeq& \left[  \frac{\mff^-_1}{\mfg^-_0}   +  \frac{\mff^-_3}{3\mfg^-_0} +  \frac{\mff^-_5}{5\mfg^-_0} + \cdots  \right]\left[1  - \left(   \frac{\mfg^-_2}{2\mfg^-_0} + \frac{\mfg^-_4}{4\mfg^-_0} +\cdots\right) + \cdots \right]  \nn\\
  &=&  \frac{M_+}3\left[ 1  + \frac{(k_0R)^2 }{15} \right]  - \frac{Z\alpha \,\rho_0}{10}  \left[ 1 +  \frac{5M_+^2}{63}  +  \frac{2(k_0R)^2}{21}  \right]   +\frac{(Z\alpha \,\rho_0)^2}{8} \left( \frac{M_+}{18} -  \frac{M_-}{50} \right) \nn\\
  && \qquad\qquad\qquad\qquad\qquad\qquad -  Z\alpha \,\rho_2 \Bigl[ 1 + \cdots \Bigr] + \cdots   \,.
\eea

\subsubsection{Uniform charge distribution}

A special case of the previous section is the case of a constant charge distribution
\be
 \rho = \frac{3Ze}{4\pi R^3} \qquad \hbox{for $r \le R$} \,,
\ee
and so represents the special case $\rho_0 = 1$ and $\rho_k = 0$ for all $k \ne 0$. For this distribution the rms radius and the moment $\langle r^3 \rangle_{(2)}$ are given explicitly by
\be
  r_p^2 = \frac{1}{Ze} \int \exd^3x \, r^2 \rho(x) = \frac{3}{R^3} \int_0^R \exd r \, r^4 = \frac{3R^2}5 \,,
\ee
and 
\bea
 \langle r^3 \rangle_{(2)} &=& \frac{1}{(Ze)^2} \int \exd^3x \,\exd^3y\; |\bfx|^3 \rho(\bfy-\bfx) \rho(\bfy) =  \frac{1}{(Ze)^2}  \int \exd^3z \, \exd^3y \; |\bfz+ \bfy|^3 \rho(\bfz) \rho(\bfy) \nn\\
 &=& \frac12 \left( \frac{3}{R^3} \right)^2 \int_0^R \exd z \int_0^R \exd y \int_{-1}^1 \exd \cos\theta \; y^2 z^2 \Bigl( y^2+z^2 + 2yz \cos\theta \Bigr)^{3/2} \nn\\
 &=& \frac1{10}\left( \frac{3}{r^3} \right)^2 \int_0^R \exd z \int_0^R \exd y\; yz\Bigl( \vert y + z \vert^5 - \vert y - z \vert^5 \Bigr) \nn\\
 &=& \frac1{5}\left( \frac{3}{r^3} \right)^2 \int_0^R \exd z\, z \left\{ \int_0^z \exd y \left[ 5y^2 z^4 + 10y^4z^2 + y^6 \right] + \int_z^R \exd y \left[ 5y^5 z + 10y^3 z^3 + y z^5 \right] \right\} \nn \\
 &=& \frac1{5}\left( \frac{3}{r^3} \right)^2 \int_0^R \exd z\, z \left[ -\frac{1}{42}z^8 + \frac{1}{2}R^2 z^6 + \frac{5}{2}R^4z^4 + \frac{5}{6}R^6 z^2 \right] = \frac{32}{21} \; R^3 \,.
\eea

The electrostatic potential coefficients for this charge distribution are
\be
  A_2 = \frac{Ze}{8\pi R} \qquad \hbox{and} \qquad  A_k = 0 \quad\hbox{for $k > 2$} \,,
\ee
and so in the continuity condition this gives   
\be
 A_0(0) + A_2   = - \frac{Ze}{4\pi R} \qquad \hbox{and so} \qquad eA_0(0) = - \frac{3Z\alpha}{2 R} \,.
\ee

The complete electrostatic potential therefore is
\be
 eA_0(r) = \frac{Z\alpha}{ R} \left[ -1 + \frac{1}{2} \Bigl( u^2 - 1 \Bigr)   \right]\,,
\ee
where $u = r/R$. Consequently
\be
  M_\pm  = mR \pm \left( \omega R + \frac{3Z\alpha}2 \right) = (m\pm W)R \pm \frac{Z\alpha}2\,,
\ee
and so
\be
 (k_0 R)^2 = - M_+M_- =  \left( WR+ \frac{Z\alpha}2 \right)^2 -(mR)^2   
 \,.
\ee

Finally, evaluating at bound-state energies $\omega \simeq m\left[1  - \frac12(Z\alpha/n)^2 + \cdots\right]$, we have $M_- \simeq -\frac32Z\alpha + \frac12 \, mR(Z\alpha/n)^2 + \cO[mR(Z\alpha)^4]$ and $M_+ \simeq  \frac32 Z\alpha+ 2mR\left[1 - (Z\alpha/2n)^2 + \cO[(Z\alpha)^4] \right] $ so their product is $(k_0R)^2 = -M_+M_- \simeq \frac94(Z\alpha)^2 + 3mRZ\alpha \left[ 1 + \cO[(Z\alpha)^2]\right]$. The boundary condition therefore becomes
\bea
  \left( \frac{g_+}{f_+} \right)_{r=R}   &=&  \frac{M_-}3\left[ 1  + \frac{(k_0R)^2 }{15} \right]  +\frac{Z\alpha }{10}  \left[ 1 +  \frac{5M_-^2}{63}  +  \frac{2(k_0R)^2}{21}  \right]   +\frac{(Z\alpha )^2}{8} \left( \frac{M_-}{18} -  \frac{M_+}{50} \right) \nn\\
  &=& - Z\alpha \left[  \frac25 + \left( \frac{107}{1400} - \frac{1}{6n^2} \right) mRZ\alpha +\frac{419}{8400}(Z\alpha)^2 + \cdots \right]  \,.
\eea

Similarly, the parity-odd expression is 
\bea
  \left( \frac{f_-}{g_-} \right)_{r=R}   &=&   \frac{M_+}3\left[ 1  + \frac{(k_0R)^2 }{15} \right]  - \frac{Z\alpha \,\rho_0}{10}  \left[ 1 +  \frac{5M_+^2}{63}  +  \frac{2(k_0R)^2}{21}  \right]   +\frac{(Z\alpha \,\rho_0)^2}{8} \left( \frac{M_+}{18} -  \frac{M_-}{50} \right)   + \cdots  \nn\\
  &=& \frac{2mR}3 + \frac{2 Z\alpha}5  + \cdots  \,.
\eea

\subsubsection*{Expansion coefficients}

For comparison, in terms of the parameters $\hat g_i$ and $\hat f_i$ found in 
\be
 \left( \frac{g_+}{f_+} \right)_{r=R} = Z\alpha \Bigl[ \hat g_1 + \hat g_2 (mRZ\alpha) + \hat g_3 (Z\alpha)^2 + \cdots \Bigr] \,,
\ee
and
\be
 \sqrt{\frac{m-\omega}{m+\omega}} \left( \frac{f_-}{g_-} \right)_{r=R} = \frac{1}{2n} \Bigl[ \hat f_1 (mRZ\alpha) + \hat f_2 (mRZ\alpha)^2 + \hat f_3 (Z\alpha)^2 + \cdots \Bigr] \,,
\ee
we have
\be
  \hat g_1 = - \, \frac25 \,, \qquad \hat g_2 =- \frac{116}{1575}  + \frac{1}{6n^2} \qquad \hbox{and} \qquad \hat g_3 = - \frac{736}{17325}  \,,
\ee
while for the parity-odd state the parameters are
\be
  \hat f_1 =  \frac23 \,, \qquad \hat f_2 = +\frac{32}{315} \qquad \hbox{and} \qquad \hat f_3 = + \frac{2}{5}  \,.
\ee

\end{document}